\documentclass[a4paper,11pt]{article}

\usepackage{authblk,amsmath,graphicx,amssymb}

\usepackage[T1]{fontenc} % if needed
\usepackage{xcolor}
\usepackage{xspace}

\usepackage{booktabs}
\usepackage{float}
\usepackage{titlesec}
\usepackage{capt-of}
\usepackage{slashed,bm}
\usepackage[font={small}]{caption}

\usepackage{array}
\usepackage{arydshln}
\usepackage[    bookmarks,
                 bookmarksopen = true,
                 bookmarksnumbered = true,
                 linktocpage,
                 colorlinks = true,
                 linkcolor = blue,
                 urlcolor  = blue,
                 citecolor = blue,
                 anchorcolor = green,
                 hyperindex = true,
                 hyperfigures]
                {hyperref}

\usepackage[sort&compress,numbers]{natbib}
\usepackage{soul}

\newcommand{\Qtop}{Q_{\rm top}}
\newcommand{\qtop}{q_{\rm top}}

\newcommand{\qtopEM}{q_{\rm top}^{\rm EM}}
\newcommand{\gayy}{g_{A\gamma\gamma}^{\text{QCD}}}
\newcommand{\gayyn}{g_{A\gamma\gamma}^{0}}
\newcommand{\dd}{\textmd{d}}
\newcommand{\Z}{\mathcal{Z}}

\newcommand{\expv}[1]{\langle#1\rangle}

\begin{document}

\title{\bf The axion-photon coupling from lattice Quantum Chromodynamics}

\author[a]{B. B. Brandt}
\author[b]{G. Endr\H{o}di}
\author[a,c]{J. J. Hern\'andez Hern\'andez}
\author[b]{G. Mark\'o}
\author[a]{L. Pannullo}
\affil[a]{Universität Bielefeld, Universitätsstraße 25, 33615 Bielefeld, Germany}
\affil[b]{Institute of Physics and Astronomy, Eötvös Lor\'and University, P\'azm\'any P. s\'et\'any 1/A, H-1117 Budapest, Hungary}
\affil[c]{Key Laboratory of Quark and Lepton Physics (MOE) and Institute of Particle Physics,
Central China Normal University, Wuhan 430079, China}

%\emailAdd{brandt@physik.uni-bielefeld.de}
%\emailAdd{gergely.endrodi@ttk.elte.hu}
%\emailAdd{hernandez@physik.uni-bielefeld.de}
%\emailAdd{gmarko@physik.uni-bielefeld.de}
%\emailAdd{lpannullo@physik.uni-bielefeld.de}
 
\maketitle

\begin{abstract}
Quantum Chromodynamics (QCD) is the theory of the strong interactions within the Standard Model of particle physics, which explains more than 99\% of the mass of the visible Universe. However, there is evidence that a substantial portion of our Universe is made up of particles beyond the Standard Model, i.e.\ dark matter. A popular dark matter candidate is the axion -- a hypothetical particle that also solves the so-called strong CP-problem, the unexpected symmetry of QCD under time reversal. 
The experimental detection of axions hinges on their conversion rate to photons, controlled by the axion-photon coupling. This coupling depends on the specific axion model, but also receives a sizable model-independent contribution from QCD. Here we present the first non-perturbative determination of the QCD contribution using continuum extrapolated lattice simulations. The calculation is based on determining the response of the QCD vacuum to time reversal-odd combinations of background electromagnetic fields. We develop two independent methods exploiting different features of this response and obtain $\gayy f_A/e^2=-0.0224(10)$ in units of the axion scale $f_A$ and the elementary charge $e$. Armed with this first-principles result, we present a novel update on how experimental observations can be used to constrain the landscape of axion models, useful for guiding contemporary and future observational strategies.
\end{abstract}

\flushbottom

\section{Introduction}

The Standard Model is the theory encompassing all known elementary particles and their interactions: the electromagnetic, weak and strong forces. It is a remarkably well tested theory that agrees with experimental measurements to extreme precision.
However, the Standard Model does not include the gravitational interaction nor dark matter. In fact, broad range of astrophysical observations, such as rotation curves of spiral galaxies~\cite{Rubin:1970zza}, velocity dispersions of galaxy clusters~\cite{Zwicky:1933gu} and weak gravitational lensing effects~\cite{Clowe:2006eq} indicate 
that besides visible matter, the Universe also contains dark matter that couples extremely weakly to the known elementary particles (see e.g.\ Ref.~\cite{Cirelli:2024ssz} for a recent review). The nature of dark matter is, as of today, unknown.

The electroweak sector of the Standard Model inherently violates time reversal (or the combination of charge conjugation and parity, CP) symmetry,
as first observed in decays of neutral kaons~\cite{Christenson:1964fg}.
In turn, the strong interactions are observed to fully respect CP symmetry, manifested through the extremely suppressed value of the neutron electric dipole moment~\cite{Abel:2020pzs}.
This is a surprising finding, because a CP-odd interaction in QCD is allowed by further symmetries of the theory -- hence it is dubbed the strong CP-problem.

Axions are hypothetical particles that may solve both of the above shortcomings. Via the Peccei-Quinn mechanism~\cite{Peccei:1977hh,Peccei:1977ur}, axions obtain a vacuum expectation value that exactly cancels possible CP-violating terms in QCD and explain the overall symmetry of the strong interactions under CP. In addition, axions are dark matter candidates that provide a possible explanation for the invisible portion of our Universe. While the microscopic details of axion physics can be described by a range of possible models, see e.g.\ Ref.~\cite{DiLuzio:2020wdo}, all of these predict a coupling $g_{A\gamma\gamma}$ between axions and photons. Together with its mass $m_A$, these constitute the most important characteristics of the axion. In particular, the axion-photon coupling provides the most promising avenue to detect axions in terrestrial experiments that search for the CP-odd electromagnetic signal that axion decays produce.

Specifically, axion models include a direct coupling $\gayyn$ to photons as well as an indirect contribution $\gayy$ through the carriers of the strong interaction, gluons. While the former depends on the model details, the latter is model-independent and is determined completely by QCD. The decay rate, relevant for experimental detection, is set by the sum $\gayyn+\gayy$ of the above two terms. Together with experimental constraints, the calculation of the QCD contribution therefore provides an effective way to test axion models and determine the properties of this, so far missing ingredient of high-energy elementary particle physics.

As of today, the only estimations of the QCD contribution to the axion-photon coupling were obtained via chiral perturbation theory (ChPT), a low-energy effective theory for the strong interactions, and
one of these estimates~\cite{GrillidiCortona:2015jxo} is used in the latest review of the Particle Data Group~\cite{ParticleDataGroup:2024cfk}.
However, different variants of ChPT~\cite{GrillidiCortona:2015jxo,Lu:2020rhp,Gao:2022xqz,Meggiolaro:2025yiu}, employing leading-order or next-to-leading order expansions, two or three quark flavors as well as effects of the axial anomaly give values in a wide range $-0.0260(4)<\gayy f_A/ e^2<-0.02064(13)$, differing from each other by several standard deviations. This calls for a non-perturbative calculation,
and in this article we present the first determination of $\gayy$ that uses solely first principles 
and does not rely on uncontrolled approximations.
Since QCD is a strongly coupled quantum field theory, we need to use computationally demanding non-perturbative lattice simulations in order to capture the full dynamics of QCD and the corresponding elementary particles, quarks and gluons. Our preliminary results were presented in Refs.~\cite{Brandt:2022jfk,Brandt:2023awt}. 

\section{The axion-photon coupling from lattice simulations}

The QCD contribution to the axion-photon coupling is a low-energy effective parameter that can be calculated by considering homogeneous background axion and electromagnetic fields~\cite{GrillidiCortona:2015jxo}. To determine $\gayy$,
we consider QCD with the three lightest quark flavors $f=u,d,s$, interacting non-perturbatively with gluons and the above classical fields. The parameters of this theory are the quark masses $m_f$ that are fixed by experimental constraints, the strong coupling $g$ 
and the quark electric charges $q_u=-2q_d=-2q_s=2e/3$ (given in units of the elementary electric charge $e$), through which quarks interact with the background electromagnetic fields. The dynamical electromagnetic interactions between quarks via photons are next-to-leading-order in the fine structure constant $\alpha=e^2/(4\pi)$, giving a correction that is much smaller than the precision of our calculation and can be safely neglected in this work.
This theory is discretised on a finite lattice with lattice spacing $a$. Our lattice setup is detailed in App.~\ref{sec:latsetup}.

A homogeneous axion field $A$ enters the Euclidean QCD action in the form $-i\Qtop A/f_A$, where $f_A$ is the axion scale parameter and $\Qtop$ the topological charge,
\begin{equation}
\Qtop=
\frac{g^2}{64\pi^2}\int\dd^4x
\,\epsilon_{\mu\nu\rho\sigma}G^a_{\mu\nu}(x)G^a_{\rho\sigma}(x)\,,
\label{eq:Qtopdef}
\end{equation}
a CP-odd topological invariant composed of
the gluon field strength $G$ that measures the winding of the fields in color space.
Electromagnetic fields $\bm E$, $\bm B$ couple to $A$ in an analogous form, with the CP-odd scalar product $\bm{EB}$
taking the place of $\Qtop$. 
With these homogeneous background fields, the coupling $\gayy$ can be obtained from the Euclidean partition function $\Z$ as~\cite{GrillidiCortona:2015jxo} \begin{equation}
    \gayy = \frac{1}{V_4}\left.\frac{\partial^2\, \log\Z}{\partial A \,\partial (\bm E \bm B)} \right|_{A=\bm E=\bm B=0}
    = \frac{1}{V_4f_A}\left.\frac{\partial \expv{\Qtop}_{\bm E\bm B}}{\partial (i\bm E \bm B)}\right|_{\bm E=\bm B=0}
    \label{eq:apc_def}
\end{equation}
where $V_4$ is the space-time volume. The differentiation with respect to the axion field results in the expectation value of the QCD topological charge in the presence of electromagnetic fields, as indicated by the subscript $\expv{.}_{\bm E\bm B}$. This expectation value vanishes at zero electromagnetic fields due to CP symmetry and develops a nonzero value when the CP-odd product $\bm E\bm B$ is nonzero.
We note that in a finite periodic volume in Euclidean space-time, the background electric field needs to be chosen purely imaginary, and moreover, both $i\bm E$ and $\bm B$ are discrete variables. Thus, to obtain $\gayy$, the differentiation in Eq.~\eqref{eq:apc_def} is to be taken numerically.\footnote{A similar approach was considered in Ref.~\cite{DElia:2012ifm} to compute the susceptibility of the QCD vacuum with respect to CP-odd electromagnetic fields.} 
The details of the calculation of $\Qtop$ on the lattice are provided in App.~\ref{app:coupl-detail}.
Analogously to $\gayy$, the mass $m_A$ of the axion in units of $f_A$ can be determined by differentiating $\log\Z$ twice with respect to $A$, leading to the topological susceptibility $\expv{\Qtop^2}_0$ of QCD, which has been determined on the lattice by various groups~\cite{Bonati:2015vqz,Borsanyi:2016ksw,Petreczky:2016vrs,Taniguchi:2016tjc,Athenodorou:2022aay}. 
We note moreover that in the combination $\gayy f_A/e^2$ the electric charge renormalisation constant cancels, rendering it a finite renormalized observable. 

We considered two methods to determine the expression~\eqref{eq:apc_def}: first, by calculating the topological charge directly in terms of the gluon fields via Eq.~\eqref{eq:Qtopdef} (referred to below as the gluonic method), and second, by expressing it indirectly via fermionic observables (fermionic method). The latter is a novel approach based on the axial Ward identity for fermionic operators, which we develop here for the first time. As we explain in App.~\ref{app:derivation_ward_identity}, evaluating
the axial Ward identity
on configurations generated with and without background electromagnetic fields allows one to express the axion-photon coupling using expectation values of the pseudoscalar operator $P_f(\bm E \bm B)$ for a given flavor $f$,
\begin{equation}
 \gayy = \lim_{\bm E \bm B\to0}\frac{1}{f_A} \frac{3q_f^2}{4\pi^2}\left[ \frac{\expv{P_f(\bm E\bm B)}_{\bm E\bm B}}{\expv{P_f(\bm E\bm B)}_{0}} - 1\right]\,,
 \label{eq:fermion1}
\end{equation}
where the expectation value in the numerator -- just as in Eq.~\eqref{eq:apc_def} -- includes the effect of electromagnetic fields on the distribution of gluon configurations, whereas it is excluded in that in the numerator.  
Eq.~\eqref{eq:fermion1} can be generalized by considering the axial Ward identity for linear combinations of quark flavors. Here we study four cases, namely one for each quark flavor independently and one for the average of the $u$ and $d$ quarks, for which the explicit expression is derived in App.~\ref{app:derivation_ward_identity}. 
We note that a possible third approach, not considered here, would be to employ the spectral projector method of Ref.~\cite{Bonanno:2019xhg} for the computation of the topological charge.

\section{Value of the axion-photon coupling and constraints on axion models}
\label{sec3}

We calculated the expectation values of the topological charge and of the pseudoscalar condensates for a range of weak electromagnetic fields on configuration ensembles with different lattice spacings. 
The details of the lattice setup and the specific parameters are included in App.~\ref{sec:latsetup}.
To evaluate $\gayy$ for either the gluonic or the fermionic methods, we need to carry out double extrapolations: $\bm E, \bm B\to0$ as well as the continuum limit $a\to0$.
We performed the $\bm E,\bm B\to0$ limit using bivariate polynomial fits, followed by univariate polynomial fits for the continuum extrapolation. 
Each of these fits were carried out in various different variants and weighted by the Akaike Information Criterion (AIC) in order to estimate the systematic uncertainties originating from different levels of operator improvement, specific extrapolation choices, contamination from higher orders in the electromagnetic fields and lattice discretisation errors. Statistical uncertainties arise from the finite sampling of the configuration ensembles and were taken into account using the jackknife procedure. The details of the extrapolation procedure are described in App.~\ref{sec:error}.

\begin{figure}[htb]
  \centering
    \includegraphics[width=\textwidth]{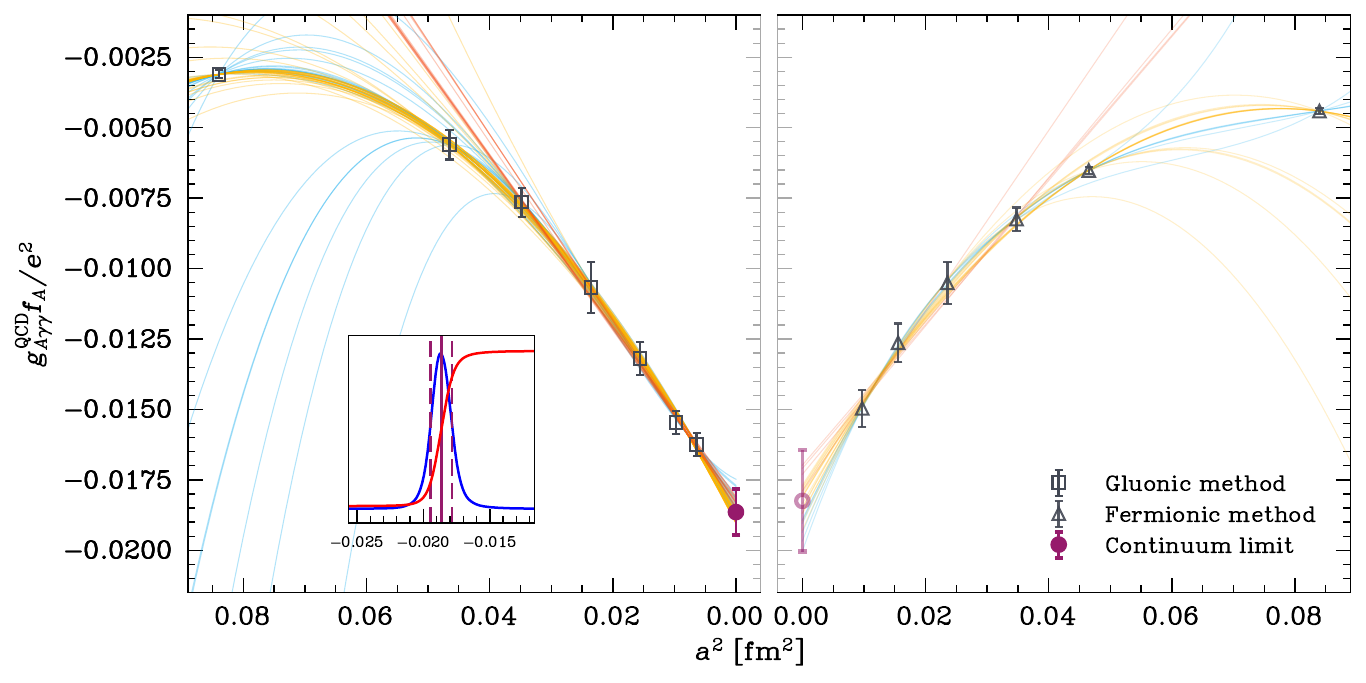}
  \caption{Continuum extrapolation of the QCD contribution to the axion-photon coupling using the gluonic method, involving an improved topological charge operator with rounding (left panel) and the fermionic method, employing the average light quark flavors (right panel). Each black data point represents the result of extrapolations to zero electromagnetic field, including statistical and systematic errors. The purple point in the left panel marks our final result, obtained by combining all extrapolations using the gluonic method. The faint purple point in the right panel corresponds to the continuum extrapolations using the fermionic method with the average light quark flavors. Polynomial fits with different orders and AIC weights are marked by different colors (red, yellow and blue curves are fits including terms up to $a^2$, $a^4$ and $a^6$, respectively) and line widths.
  The inset depicts the probability density and the cumulative density of the final result. The vertical purple line is the median of the distribution and the dashed lines mark the range corresponding to the total error.
  }
  \label{fig:clim}
\end{figure}

Representative fits for the continuum extrapolation are shown in Fig.~\ref{fig:clim} for the gluonic and fermionic methods, demonstrating that 
both give consistent results in the continuum limit.
Our most accurate determination stems from the gluonic method, and the final result that we obtain for the axion-photon coupling is
$-0.0186(8) \cdot e^2/f_A$.
This value corresponds to the case of mass-degenerate up and down quarks. As explained in App.~\ref{app:isospin}, the effect of the mass splitting between the light quarks gives rise to a correction of about $20\%$, 
so that the final physical value is
\begin{equation}
    \gayy = -0.0224(10) \cdot \frac{e^2}{f_A}
= -1.77(8)\cdot \frac{\alpha}{2\pi f_A}
= -3.92(27) \,\frac{e^2m_A}{\text{GeV}^2}\,,
\label{eq:finalres}
\end{equation}
where we also expressed the result using the fine structure constant $\alpha$ in order to facilitate comparisons with the literature. The error budget of the final result is 
\begin{equation}
    \gayy f_A/e^2=-0.0224(2)_{\rm stat}(2)_{\rm def}(5)_{\rm a}(8)_{\rm EB}(1)_{\rm vol}(2)_{\rm m},
\end{equation}
where the first bracket indicates the statistical uncertainty, the second the systematic error associated to the different topological charge definitions, the third and fourth the systematic errors corresponding to the continuum and $\bm E,\bm B\to0$ extrapolations, the fifth to the volume effects and the last to the mass correction factor. The details of the individual terms are discussed in App.~\ref{sec:error}.
This result corresponds to a reduction by about $10\%$ of the magnitude of the coupling compared to the prediction by next-to-leading-order two-flavor chiral perturbation theory~\cite{GrillidiCortona:2015jxo}, quoted in the PDG~\cite{ParticleDataGroup:2024cfk}, $\gayy = -1.92(4)\cdot \alpha/(2\pi f_A)$. We note moreover that among the chiral perturbation theory predictions, the one taking into account the effects of the axial anomaly~\cite{Meggiolaro:2025yiu} lies the closest to our result.
In~\eqref{eq:finalres} we also expressed the coupling in units of the axion mass, determined from the topological susceptibility~\cite{Borsanyi:2016ksw}, which is free of the unknown scale $f_A$.
This is the first determination of this combination of axion parameters using only first-principle methods. 

Our pioneering non-perturbative calculation of the QCD contribution $\gayy$ to the axion-photon coupling can be used to update existing constraints on the axion parameter landscape, often used in the context of observational experiments, with first principles input. Furthermore, we demonstrate below how the knowledge of the precise non-perturbative value of $\gayy$ can directly narrow down the set of experimentally allowed axion models.
Experiments searching for axion signals are sensitive to the magnitude of the total coupling $g_{A\gamma\gamma}=\gayyn+\gayy$, composed of the model-dependent term and the QCD contribution we calculated. The former can be written using the color ($N$) and electromagnetic ($E$) anomaly coefficients specific to the realisation of the Peccei-Quinn symmetry in the model,  $\gayyn=E/N\cdot \alpha/(2\pi f_A)$. In principle the total coupling is subject to renormalisation group effects, however we only need the value of the coupling below the typical QCD scale $\sim 1$ GeV, where these effects are generally considered negligible~\cite{Bauer:2017ris,Bauer:2020jbp,Choi:2021kuy,DiLuzio:2023tqe}. The phenomenologically preferred axion models correspond to the range $0\le E/N\le 44/3$ according to e.g.\ Ref.~\cite{DiLuzio:2016sbl}.\footnote{For less restrictive selection criteria, the upper bound increases to $E/N=170/3$~\cite{DiLuzio:2016sbl}.} Among these, the most widely known models include the Kim-Shifman-Vainshtein-Zakharov (KSVZ)~\cite{Kim:1979if,Shifman:1979if} ($E/N=0$) and the Dine-Fischler-Srednicki-Zhitnitsky (DFSZ)~\cite{Zhitnitsky:1980tq,Dine:1981rt} ($E/N=8/3$) models.

\begin{figure}[htb]
  \centering
    \includegraphics[width=\textwidth]{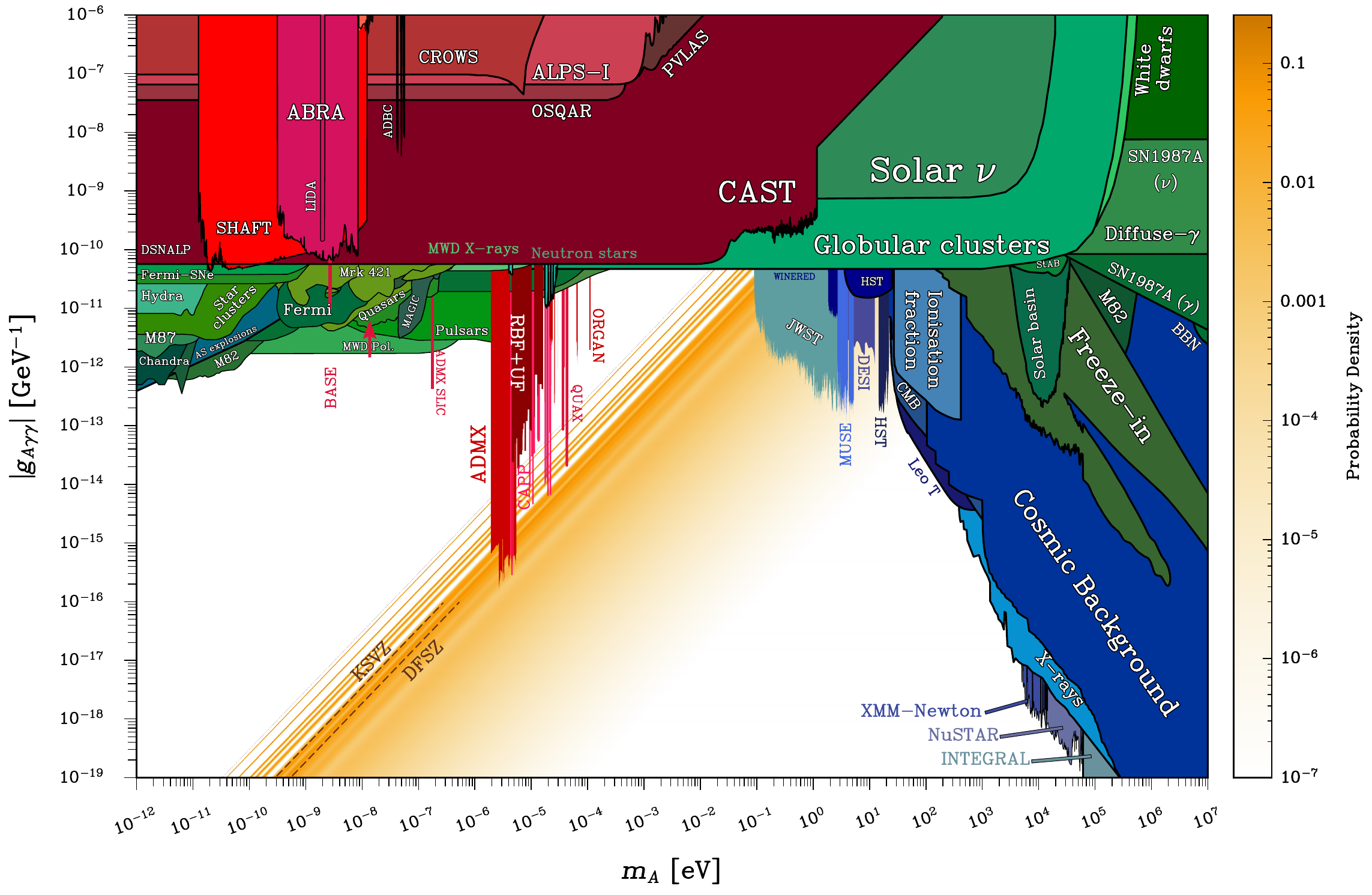}
  \caption{Experimental constraints on the axion parameter landscape, adapted from Ref.~\cite{AxionLimits}. The probability density of the magnitude of the total axion-photon coupling, given our result for the probability distribution of $\gayy$ from Fig.~\ref{fig:clim} and the preferred models of Ref.~\cite{DiLuzio:2016sbl}, is visualized by the orange shaded region using a logarithmic color map. 
  The KSVZ and DFSZ models are highlighted by the dashed lines.
  For details on the experimental data, see Refs.~\cite{AxionLimits,ParticleDataGroup:2024cfk} and references within. 
}
  \label{axion-photon-full}
\end{figure}

The axion parameter landscape is traditionally represented in the $|g_{A\gamma\gamma}|-m_A$ plane with logarithmic scales, as shown in Fig.~\ref{axion-photon-full}. 
Various regions of this plot are ruled out by different experiments, as visualized by the colored areas. 
Setting the fine structure constant to its value $\alpha=1/137$ in the Thomson limit, given values of the axion parameters in units of $f_A$ correspond to straight lines in the figure. 
Our result for $\gayy$, together with the $\gayyn$ values of the preferred axion models produces a probability density for such lines, see App.~\ref{sec:error} for details. This probability density is visualized in Fig.~\ref{axion-photon-full} by the orange color map, covering the region of possible axion parameters.
Specifically, the upper bound for $|g_{A\gamma\gamma}|$ corresponds to the model with $E/N = 44/3$ and the lower one to that with $E/N=5/3$.\footnote{Notice that the same models give the extremal values for the two-flavor next-to-leading-order chiral perturbation theory prediction~\cite{GrillidiCortona:2015jxo} as well.} For the latter model, the total coupling is compatible with zero within two standard deviations, leading to
the nonzero values of the probability density towards the bottom of this logarithmic plot.
The results for the KSVZ ($g_{A\gamma\gamma}f_A/\alpha =-0.282(13)$) and DFSZ models ($0.143(13)$) are also separately indicated in the figure. We highlight that for the majority of the models, $\gayy$ is of similar magnitude as the model-dependent direct coupling $\gayyn$, but of opposite sign. The QCD contribution is thus crucial for predicting the axion-to-photon conversion rate for experiments.

\begin{figure}[htb]
  \centering
    \includegraphics[width=\textwidth]{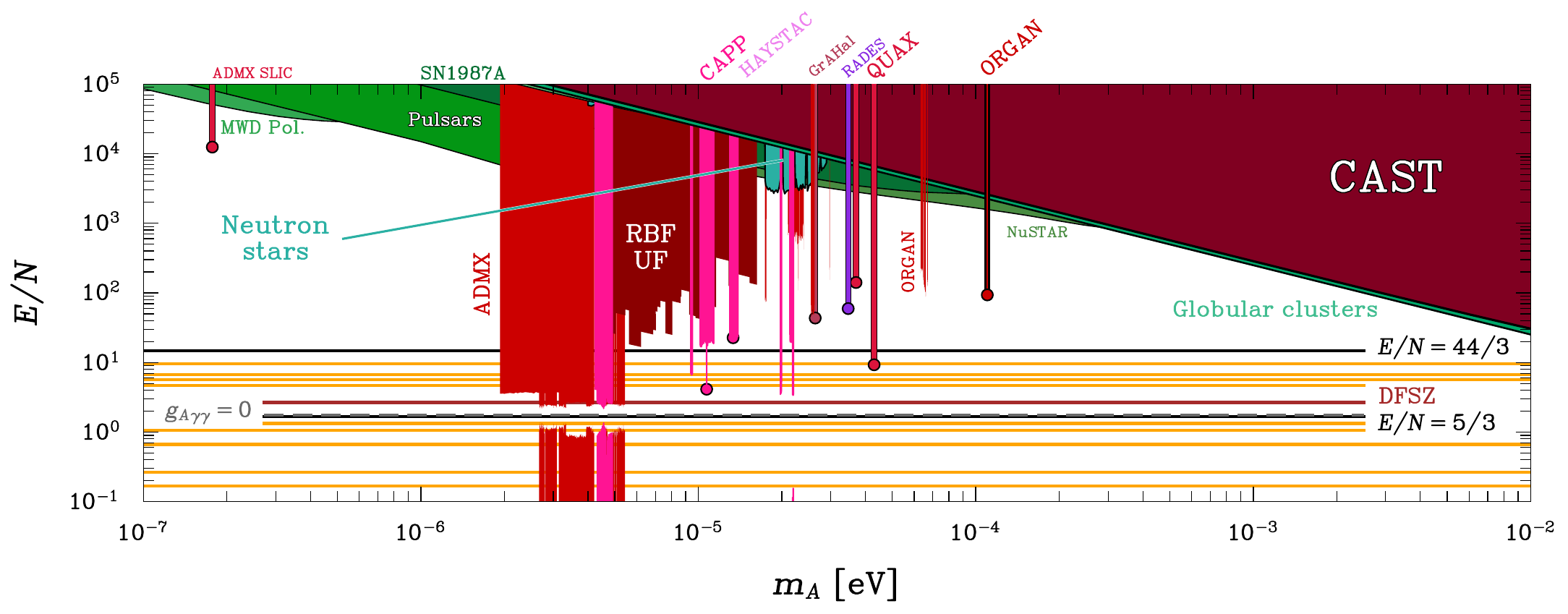}
  \caption{Constraints on the experimentally allowed values of the model-dependent contribution to the axion-photon coupling, adapted from Ref.~\cite{AxionLimits}. For details on the experimental data, see Refs.~\cite{AxionLimits,ParticleDataGroup:2024cfk} and references within. 
The black lines denote the models which give the biggest and smallest values for $|g_{A\gamma\gamma}|$. The brown line marks the DFSZ model with $E/N = 8/3$ and the gray dashed line the value where the total coupling vanishes. Finally, the orange lines correspond to the rest of the models considered in Ref.~\cite{DiLuzio:2016sbl}. Some of the lines have been cut horizontally for visualization purposes.
}
  \label{axion-photon-models}
\end{figure}

Fig.~\ref{axion-photon-models} shows the possible values of $E/N$ given the experimental constraints and our value for the QCD contribution.
Specifically, the experimental constraints have been obtained by solving directly for $E/N$, taking into account the uncertainty of $\gayy$. 
The upper and lower bounds mentioned in the previous paragraph, together with the negative of our result for the QCD contribution are also indicated in the figure. Axion models with $E/N\approx1.77$ would correspond to a vanishing total coupling $g_{A\gamma\gamma}\approx0$ and would therefore be practically impossible to detect through this channel. As mentioned above, among the discussed models~\cite{DiLuzio:2016sbl}, the closest to this value is the model with $E/N=5/3$. 
We note that while the experiments like CAST and ADMX were originally devised in the context of the minimal models as KSVZ and DFSZ, the corresponding bounds can also be interpreted in terms of more general axion models. 
Furthermore, assuming that 1 to 50\% of the amount of dark matter is made up of axions arising from the misalignment mechanism in a post-inflation scenario, a lattice QCD estimate for $m_A$ lies in the range $50\:\mu\text{eV}\,<m_A<\,1500\:\mu\text{eV}$~\cite{Borsanyi:2016ksw}.
We highlight that the region obtained from this axion mass determination and our calculation of the axion-photon coupling has still not been explored by current experiments.
Future experiments \cite{IAXO:2020wwp,Stern:2016bbw,Alesini:2023qed,Bajjali:2023uis,BREAD:2021tpx,Beurthey:2020yuq,ALPHA:2022rxj,Ouellet:2018beu,Salemi:2021gck,Crisosto:2019fcj,Gramolin:2020ict,Zhang:2021bpa,DMRadio:2022pkf,Aybas:2021nvn} are, however, planned to focus on different regions of the yellow band in Fig.~\ref{axion-photon-full} (see \cite{Ringwald:2024uds} for an overview). This again underlines the importance of knowing the precise value of the model-independent part of the coupling, determined purely from first-principles QCD.

\section*{Acknowledgments}

The authors acknowledge support by the Deutsche Forschungsgemeinschaft (DFG, German Research Foundation) through the CRC-TR 211 `Strong-interaction matter under extreme conditions' -- project number 315477589 -- TRR 211, by the Helmholtz Graduate School for Hadron and Ion Research (HGS-HIRe for FAIR), the Hungarian National Research, Development and Innovation Office - NKFIH (Research Grant Hungary 150241) and the European Research Council (Consolidator Grant 101125637 CoStaMM).
Views and opinions expressed are however those of the authors only and do not necessarily reflect those of the European Union or the European Research Council. Neither the European Union nor the granting authority can be held responsible for them.
JH also acknowledges partial funding by the National Natural Science Foundation of China under Grants No. 12293064, No. 12293060, and No. 12325508.
The authors are grateful to Ahmed Ayad, Eduardo Garnacho, S\'andor Katz and Guy D.\ Moore for enlightening discussions. Part of the computations in this work were performed on the GPU cluster at Bielefeld University. We also acknowledge the Digital Government Development and Project Management Ltd.\ for awarding us access to the Komondor HPC facility based in Hungary.

\appendix

\section{Lattice setup and parameters}
\label{sec:latsetup}

We have simulated QCD with three flavors of dynamical quarks in the presence of background electromagnetic fields. Using the rooted staggered fermion discretisation, the partition function $\Z$ can be written using the Euclidean path integral over gluon links $U$ as,
\begin{equation}
    \Z = \int \mathcal{D}U e^{-\beta S_g[U]}\prod_{f=u,d,s} \left\{ \det \left[\slashed{D}_f(U,q_fE,q_fB) + m_f\right]\right\} ^{1/4},
    \label{eq:Z}
\end{equation}
where the product runs over the up, down and strange quarks and $\beta = 6/g^2$ denotes the inverse gauge coupling.
To sample the path integral~\eqref{eq:Z}, we generated ensembles with seven different values of $\beta$, corresponding to lattice spacings ranging down to $a=0.08\textmd{ fm}$. Our lattice geometries are $N_s^3\times N_t$, where $N_s$ and $N_t$ denote the number of points in the spatial and temporal directions, respectively. This setting corresponds to temperatures $T=1/(N_ta)$ that are much smaller than the typical QCD scale and approximately constant volumes $V=L^3=(aN_s)^3$. The four-dimensional volume is denoted by $V_4=V/T$. For all ensembles, the linear size of the system satisfies $m_\pi L\gtrsim 3.2$ in units of the mass of the lightest excitation, the pion. In addition, finite volume and nonzero temperature effects are explicitly checked to be smaller than our statistical errors on one of our ensembles, and are included in our final error budget. The ensemble sizes and spacings are tabulated in Tab.~\ref{tab:1}.

\begin{table*}[htb] \centering
\begin{small}
\begin{tabular}{ccccccc}
\toprule
$24^3\times32$ & $24^3\times32$ & $32^3\times48$ & $32^3\times48$ & $40^3\times48$ & $48^3\times64$ & $60^3\times60$\\
$\beta = 3.45$ & $\beta = 3.55$ & $\beta = 3.6$ & $\beta = 3.67$ & $\beta = 3.75$ & $\beta = 3.85$ & $\beta=3.95$ \\
$ a = 0.29\, \text{fm}$ & $ a = 0.22\, \text{fm}$ & $ a = 0.19\, \text{fm}$ & $ a = 0.15\, \text{fm}$ & $ a = 0.12\, \text{fm}$ & $ a = 0.10\, \text{fm}$ & $ a = 0.08\, \text{fm}$\\

\begin{tabular}{cc}
\toprule
$n_e$ & $n_b$ \\ \midrule
0 & 0 \\ \hdashline
3 & 8 \\ \hdashline
4 & 12 \\ \hdashline
4 & 18 \\ \hdashline
6 & 16 \\ \hdashline
6 & 20 \\ \hdashline
8 & 18 \\ \hdashline
8 & 21 \\
%\bottomrule
\end{tabular}&

\begin{tabular}{cc}
\toprule
$n_e$ & $n_b$ \\ \midrule
0 & 0 \\ \hdashline
3 & 3 \\ \hdashline
3 & 6 \\ \hdashline
3 & 9 \\ \hdashline
4 & 9 \\ \hdashline
5 & 9 \\ \hdashline
- & - \\ \hdashline
- & - \\
%\bottomrule
\end{tabular}&

\begin{tabular}{cc}
\toprule
$n_e$ & $n_b$ \\ \midrule
0 & 0 \\ \hdashline
2 & 7 \\ \hdashline
4 & 7 \\ \hdashline
6 & 7 \\ \hdashline
8 & 7 \\ \hdashline
7 & 10 \\ \hdashline
7 & 12 \\ \hdashline
- & - \\
%\bottomrule
\end{tabular}&

\begin{tabular}{cc}
\toprule
$n_e$ & $n_b$ \\ \midrule
0 & 0 \\ \hdashline
2 & 3 \\ \hdashline
3 & 3 \\ \hdashline
3 & 6 \\ \hdashline
4 & 6 \\ \hdashline
3 & 10 \\ \hdashline
4 & 9 \\ \hdashline
6 & 7 \\
%\bottomrule
\end{tabular}&

\begin{tabular}{cc}
\toprule
$n_e$ & $n_b$ \\ \midrule
0 & 0 \\ \hdashline
2 & 2 \\ \hdashline
1 & 7 \\ \hdashline
2 & 7 \\ \hdashline
4 & 4 \\ \hdashline
7 & 3 \\ \hdashline
4 & 7 \\ \hdashline
5 & 7 \\ 
%\bottomrule
\end{tabular}&

\begin{tabular}{cc}
\toprule
$n_e$ & $n_b$ \\ \midrule
0 & 0 \\ \hdashline
2 & 3 \\ \hdashline
3 & 4 \\ \hdashline
2 & 9 \\ \hdashline
4 & 6 \\ \hdashline
6 & 5 \\  \hdashline
- & - \\ \hdashline
- & - \\
%\bottomrule
\end{tabular}&

\begin{tabular}{cc}
\toprule
$n_e$ & $n_b$ \\ \midrule
5 & 5 \\ \hdashline
- & - \\ \hdashline
- & - \\ \hdashline
- & - \\ \hdashline
- & - \\ \hdashline
- & - \\  \hdashline
- & - \\ \hdashline
- & - \\
%\bottomrule
\end{tabular}\\

\bottomrule
\end{tabular}
\caption{\label{tab:1}Parameters of the simulations in our study. 
For each ensemble, we have generated several hundred configurations, separated from each other by five molecular dynamics trajectories.}
\end{small}
\end{table*}

In Eq.~\eqref{eq:Z}, $S_g$ is the tree-level Symanzik improved gluon action and $\slashed{D}_f$ the staggered Dirac operator with twice stout-smeared gluon links. The Dirac operator depends on the quark flavor through the quark electric charges.
In the simulations we use degenerate up and down quark masses, tuned together with the strange quark mass to the physical point 
using the line of constant physics determined in Ref.~\cite{Borsanyi:2010cj}. The $1/4$ power of the fermion determinant in Eq.~\eqref{eq:Z} arises due the rooting procedure, see e.g.\ Ref.~\cite{Durr:2005ax}.

For the determination of the low-energy effective coupling between the homogeneous axion field and the electromagnetic fields, the latter can be considered as classical background fields in the Dirac operator. Corrections to $\gayy$ due to dynamical photon interactions arise at the order $\mathcal{O}(\alpha^2)$ and are therefore expected to modify our result by about one percent and can be safely neglected. The final observable~\eqref{eq:apc_def} is defined via a slope at vanishing electric field, which we approach through simulations at imaginary electric fields, avoiding the complex action problem present for real ones~\cite{Endrodi:2024cqn}.
The electromagnetic potential $A_{\mu}$ was included in a similar fashion as the gluonic one, entering the Dirac operator as $\mathrm{U}(1)$ phases $u_{\mu,f}=\exp{(iaq_fA_{\mu})}$ that multiply the $\mathrm{SU}(3)$ gluonic links. The potential was chosen in such a way that it creates homogeneous (imaginary) electric and magnetic fields pointing in the positive $z$ direction. Due to finite lattice sizes and the periodic boundary conditions for the links, both the electric and magnetic field are quantised: $eE = 6\pi n_e/(a^2N_sN_t)$, $eB = 6\pi n_b/(aN_s)^2$~\cite{Bali:2011qj}. The flux quanta $n_e, n_b \in \mathbb{Z}$ are also listed in Tab.~\ref{tab:1}. To carry out the continuum extrapolation for the gluonic (fermionic) method, we used seven (six) lattice spacings.

We note that alternative approaches to introduce weak electromagnetic fields also exist in the literature, see the recent review~\cite{Endrodi:2024cqn}.
For the observable at hand, these approaches yield topology-current or topology-current-current correlators, which were found to require vastly higher statistics~\cite{Brandt:2022jfk}. This prompted us to focus on calculating expectation values directly at nonzero electromagnetic fields.
For the fermionic method, expectation values evaluated on $\bm E=\bm B=0$ ensembles are also required, see Eq.~\eqref{eq:fermion1}. Here we reused existing configurations from previous studies~\cite{Bali:2011qj,Bali:2012zg} too. 

\begin{figure}[htb]
  \centering
    \includegraphics[width=.5\textwidth]{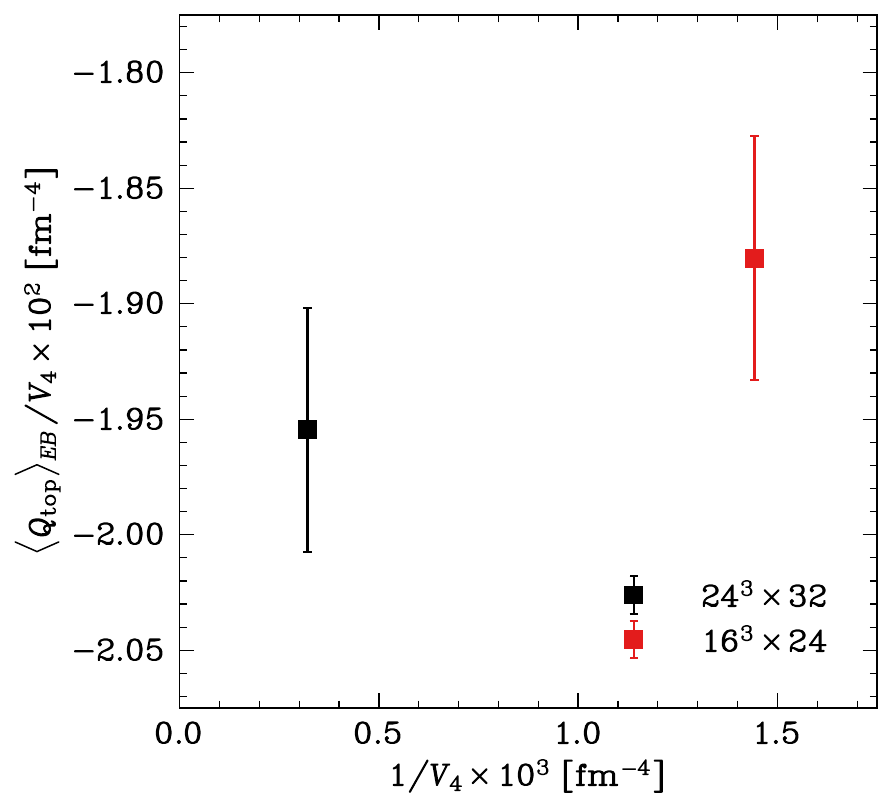}
  \caption{Comparison of the expectation value of the topological charge obtained on two different volumes (realizing slightly different low temperatures) at the same value of $\bm E\bm B$ 
  and of the lattice spacing. We observe an agreement between the two setups within errors.}
  \label{vol-temp-effects}
\end{figure}

To assess the impact of finite size and nonzero temperature effects, 
we compared the results on our $24^3\times32$ lattice with $\beta=3.45$ and $n_e = 4$, $n_b=18$ to simulations on $16^3\times24$ at the same lattice spacing and with $n_e=2$, $n_b=8$. These choices realize the same electromagnetic fields in physical units but two different values for the physical volume and for the (low) temperature. 
The results for $\expv{\Qtop}_{\bm E\bm B}$ are shown in Fig.~\ref{vol-temp-effects}, revealing an agreement between the two lattices within statistical errors. The slight downward trend towards $V_4\to\infty$ amounts to an effect of about $\approx0.0001$ in $\gayy f_A/e^2$, which is considerably smaller than other uncertainties and is included in our final error.

\section{Analysis details}
\label{app:coupl-detail}

Here we include the details of the gluonic and fermionic methods, used to calculate the axion-photon coupling on a finite lattice at nonzero background electromagnetic fields. In addition, we discuss the impact of the difference between the up and down quark masses on our final results.

\subsection{Discretisation of the topological charge and gradient flow}
\label{app:topchargedetail}

The fundamental observable for the gluonic method of the computation of $\gayy$ is the topological charge density operator $\qtop(x)$, defined as the integrand of Eq.~\eqref{eq:Qtopdef}.
In the continuum, the spatial integral $\Qtop$ of this operator is associated with the degree of winding of the gluon field in group space and assumes integer values owing to the Atiyah-Singer index theorem~\cite{Atiyah:1968mp}. 
In turn, on the lattice, the operator $\qtop$ is well known to be subject to discretisation errors due to ultraviolet fluctuations of the gauge fields, which shift the topological charge away from its continuum values, i.e.\ integers. These lattice artifacts are largely controlled by the exact form of the operator:
the simplest discretisation of $\qtop(x)$ uses the smallest closed gluonic loops (the plaquettes) for the construction, leading to lattice artifacts of $\mathcal{O}(a^2)$ at tree level. In turn, an improved version of the operator has been proposed in Ref.~\cite{Bilson-Thompson:2002xlt}, which employs larger loops, so that leading artifacts in the operator are of $\mathcal{O}(a^4)$ at tree level.
We will refer to the former definition as ``regular'' and to the latter as ``improved''.
Besides this kind of improvement, ultraviolet fluctuations can be further reduced by using the gradient flow~\cite{Luscher:2010iy} of the gauge fields,
which we define in terms of the Wilson action. The gluon links evolved to non-zero flow time $\tau_f$ are used to construct the regular or improved definitions of the discretised topological charge operators. 

Wilson flow can be effectively seen as an averaging of the gauge fields over a domain with mean-squared radius $\sqrt{8\tau_f}$, thus suppressing ultraviolet fluctuations. In addition, observables defined at non-zero flow time are renormalized and remain finite in the continuum limit~\cite{Luscher:2013cpa}. A (small) finite amount of flow thus facilitates the extraction of the topological structure of a configuration and leads to topological charge values which are close to integers, as we will demonstrate below. However, an excessive amount of flow destroys the topological information encoded in the gauge field. At finite temperatures
it is conventional to limit the flow time to $\tau_f^{\rm max} = 1/\left(8T^2\right)$~\cite{Petreczky:2016vrs}. For our study at vanishing temperature, we employed a maximal flow time of $\tau_f^{\rm max} = 1\textmd{ fm}^2$, fixed in physical units.
Below this upper bound and for a sufficient amount of flow, the topological charge is expected to be independent of the flow time and develops a plateau from which $\Qtop$ can be read off. In our study we 
have ensured that such a plateau indeed develops. 

\begin{figure}[ht]
  \centering
    \begin{minipage}[ht]{0.49\textwidth}
    \includegraphics[width=\textwidth]{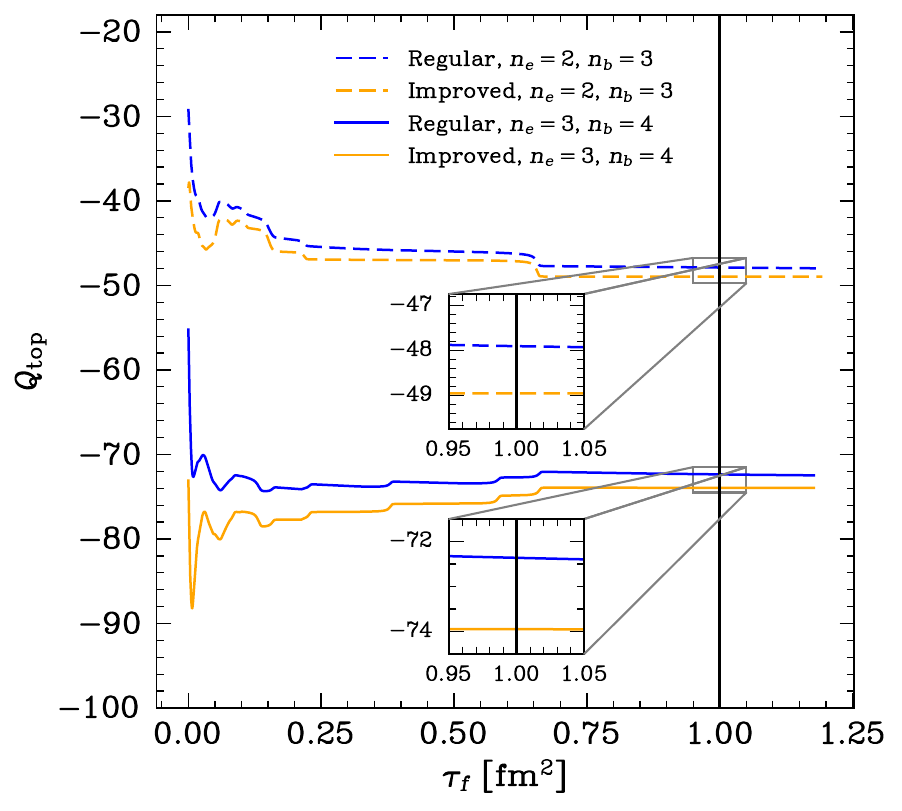}
  \end{minipage}
  \begin{minipage}[ht]{0.49\textwidth}
    \includegraphics[width=\textwidth]{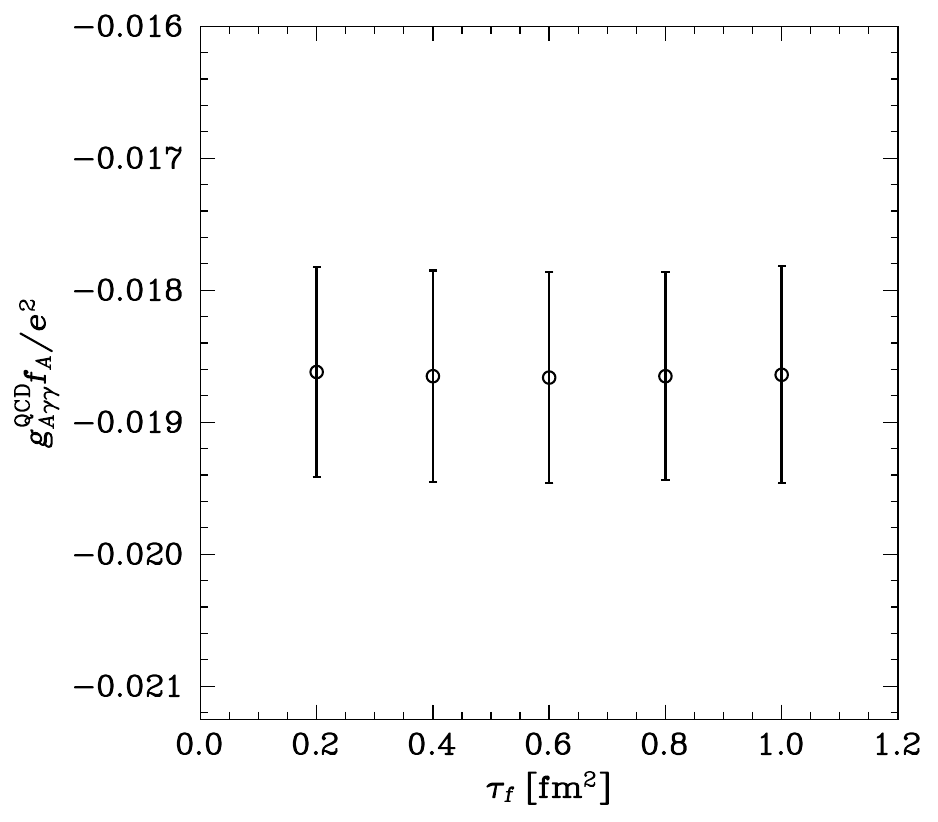}
  \end{minipage}
  \caption{Left panel: evolution of the regular and improved definitions of the topological charge with the gradient flow for two individual $48^3\times 64$ configurations and two different electromagnetic field fluxes ($n_e=2$, $n_b=3$; $n_e=3$, $n_b=4$). The maximal flow time value ($\tau_f = 1$ fm$^2$) is indicated by the dark vertical line. In this region, the topological charge has already reached a plateau.
  Right panel: continuum extrapolated result for $\gayy$ as a function of the employed flow time $\tau_f$.}
  \label{conf_flow}
\end{figure}

To demonstrate the efficacy of the gradient flow technique in the presence of the background fields, we show the $\tau_f$-dependence of $\Qtop$ on two individual $48^3\times 64$ configurations for two different electric and magnetic field flux combinations in Fig.~\ref{conf_flow}. For the integration of the flow equation we have used an adaptive step-size Runge-Kutta 3 method, implemented in the \texttt{SIMULATeQCD} framework~\cite{Mazur:2021zgi,HotQCD:2023ghu}. As can also be seen in the plots, this method leads to stable plateaus and almost integer values for the topological charge of the individual configurations appearing in ensembles close enough to the continuum limit.

To further scrutinize lattice artifacts,
we also perform an analysis with $\Qtop$ rounded to the nearest integer for each configuration, after computing it with the method described above. Furthermore, we vary the amount of flow used in the definition of the links and use the alternative flow time value $0.8\,\tau_f^{\rm max}$.
In our analysis, we studied three of these definitions (regular, improved and improved with rounding) at both $\tau_f = \tau_f^{\rm max}$ and $0.8\,\tau_f^{\rm max}$ in order to assess the influence of the choice of discretisation and amount of flow time on the observable. In the right panel of Fig.~\ref{conf_flow}, the continuum extrapolated value of $\gayy$ is shown as a function of $\tau_f$ in a broad range. The plot demonstrates the insensitivity of the continuum limit to the gradient flow depth. The systematics originating from this small effect are included in our error budget.

The use of the staggered discretisation of the Dirac operator for the sampling of the path integral may lead to another type of lattice artifact, originating from the absence of exact topological zero modes, see Refs.~\cite{Borsanyi:2016ksw,Brandt:2024gso}.
These can be mitigated in principle via an appropriate reweighting procedure, as demonstrated for the topological susceptibility $\expv{\Qtop^2}_0$ at high temperature in Ref.~\cite{Borsanyi:2016ksw}. In turn,
at zero temperature the mixing of the topological would-be zero modes and the close-to-zero modes associated with the non-vanishing chiral condensate impede a direct application of this method, as discussed in Ref.~\cite{Brandt:2024gso}. However, while for the topological susceptibility lattice artifacts at, for example $a\approx0.1\textmd{ fm}$, exceed the continuum value by an order of magnitude~\cite{Borsanyi:2016ksw}, for $\gayy$ they constitute an effect of about $20\%$. Therefore, in this work we do not employ reweighting methods of this type.\footnote{We note furthermore that at $\bm E\bm B\neq0$ the total number of topological zero modes is different for fermion flavors with different electric charges. This leads to a competition between reweighting factors for each flavor and an overlap problem that complicates reweighting methods in comparison to the situation in the absence of electromagnetic fields.}

\subsection{Fermionic definition}
\label{app:fermiondef}

For the fermionic method, discussed in detail in appendix~\ref{app:derivation_ward_identity} below, the central operator is the pseudoscalar condensate, which we computed using Wick's theorem as
\begin{equation}
    P_f(\bm E\bm B) = \frac{1}{4}\text{Tr}\left[\Gamma_5 (\slashed{D}_f+m_f)^{-1}\right]\,.
    \label{eq:trace}
\end{equation}
Here, $\Gamma_5$ is the staggered equivalent\footnote{The discretisation of $\gamma_5$ in the staggered formulation is the taste singlet matrix $\Gamma_5$, see the definition in e.g.\ Refs.~\cite{Durr:2013gp,Brandt:2024wlw}.} of $\gamma_5$, the factor $1/4$ originates from the rooting procedure and the trace is implied over color and spatial indices. Moreover, we highlighted that the operator $P_f$ depends explicitly on the electromagnetic fields through the Dirac operator. In order to evaluate Eq.~\eqref{eq:trace}, we have employed noisy estimators sampled from a Gaussian distribution. Typically, we used 100 vectors for extracting the observable, and on the $a=0.1\textmd{ fm}$ ensemble we increased this number to 150. % for noise reduction.

\begin{figure}[htb]
  \centering
  \includegraphics[width=0.5\textwidth]{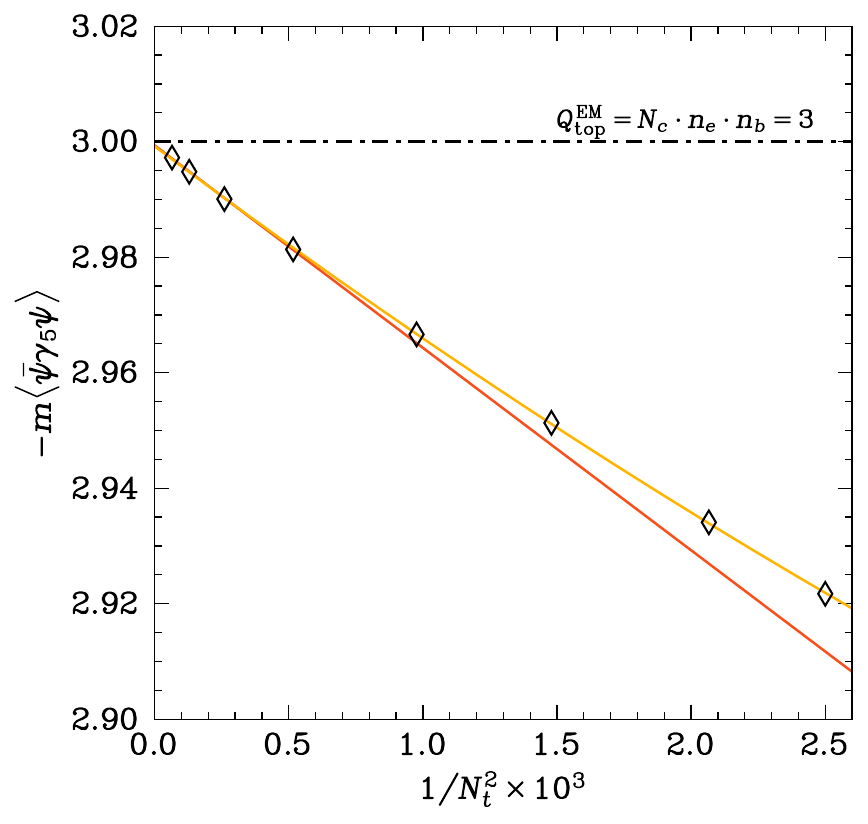}
  \caption{Check of the axial Ward identity in the free case, with $m/T = 1$, $LT = 1$ and $n_e=n_b=1$. The continuum limit is carried out in two ways, using a linear (red) and a quadratic (orange) fit in $N_t^{-2}$, extrapolating to the expected value $\Qtop^{\rm EM} = 3$.}
  \label{free_case}
\end{figure}

In order to test the behavior of the pseudoscalar condensate with the electromagnetic field, we carried out 
a benchmark in the absence of gluonic interactions, which we show in Fig.~\ref{free_case}. In this case, all gluon links equal the unit matrix, thus $\Qtop=0$ and only the electromagnetic topology contributes to the axial Ward identity, Eq.~\eqref{eq:awi} below.
For these calculations, we considered a symmetric Euclidean box $LT=1$, a single quark flavor with mass $m/T = 1$ and flux quanta $n_e=n_b=1$. To approach the continuum limit, we used temporal extents of $N_t = 20$, $22$, $26$, $32$, $44$, $62$, $88$ and $124$.

In the absence of gluonic interactions, the squared Dirac operator can be decomposed into two two-dimensional blocks (one for the magnetic field and one for the imaginary electric field) and diagonalised exactly, similarly to Ref.~\cite{Brandt:2024wlw}.
Our investigations confirm, as can be seen in Fig.~\ref{free_case}, that the axial Ward identity~\eqref{eq:awi1} is fulfilled in the continuum limit. The results at nonzero $a$ reveal that quarks experience an effective electromagnetic field that is slightly smaller in magnitude than the one built into the simulations. This effect cancels in the ratio~\eqref{eq:fermion1} used in the fermionic method, suggesting that the latter exhibits smaller discretisation effects than the gluonic method.

\subsection{Isospin breaking effects}
\label{app:isospin}

Our simulations were performed with mass-degenerate light quarks. In nature, there is a small splitting between the up and down quark masses. For most QCD observables, such isospin symmetry breaking effects amount to sub-percent corrections, like for the proton-neutron mass difference~\cite{BMW:2014pzb} or the anomalous magnetic moment of the muon~\cite{Borsanyi:2020mff}. However, observables of topological nature may behave differently in this respect. For example, the topological susceptibility is known to exhibit isospin breaking effects of about $12\%$~\cite{Borsanyi:2016ksw}, which, however, can be corrected using chiral perturbation theory. The situation for the axion-photon coupling is very similar and we follow the same strategy as Ref.~\cite{Borsanyi:2016ksw} for the topological susceptibility. In particular, within leading-order chiral perturbation theory, the ratio of axion-photon couplings at the physical and mass-degenerate points reads~\cite{GrillidiCortona:2015jxo}, 
\begin{equation}
    \frac{g_{A\gamma\gamma}^{\rm QCD, \, phys}}{g_{A\gamma\gamma}^{\rm QCD, \, sym}} = \frac{2}{5}\frac{m_u+4m_d}{m_u+m_d} = 1.208(10),
    \label{eq:factor}
\end{equation}
which shows that the small difference between the up and the down quark masses\footnote{The quark mass values were taken from Eq.~(41) of the latest FLAG review~\cite{FlavourLatticeAveragingGroupFLAG:2024oxs}.} generates a $20\%$ increase in the magnitude of the coupling with respect to the mass-degenerate case. 

We highlight that -- in contrast to the actual values $g_{A\gamma\gamma}^{\rm QCD, \, phys}$ and $g_{A\gamma\gamma}^{\rm QCD, \, sym}$, which fluctuate considerably between two- and three-flavor ChPT at leading and next-to-leading orders -- the correction factor agrees with Eq.~\eqref{eq:factor} for all existing variants of the low-energy theory. The specific values were calculated based on Refs.~\cite{GrillidiCortona:2015jxo,Lu:2020rhp,Meggiolaro:2025yiu} and listed in Tab.~\ref{tab:2}. 
Therefore, to obtain the final result for the axion-photon coupling at the physical point, we can safely rescale our determination at the mass-degenerate point by one of these predictions, and in particular we considered the $\mathrm{SU}(2)$ NLO value. 

\begin{table}[ht!]
\centering
\begin{tabular}{ccccc}
\toprule
SU(2) LO  & SU(3) LO  & SU(2) NLO & SU(3) NLO & WDV \\
1.208(10) & 1.208(10) & 1.202(10) & 1.200(10) & 1.214(11) \\ \bottomrule
\end{tabular}
\caption{Ratio of the axion-photon couplings at the physical point and at the mass-degenerate point for different variants of ChPT~\cite{GrillidiCortona:2015jxo,Lu:2020rhp,Meggiolaro:2025yiu}. Notice that all predictions agree within errors, signaling an identical mass-dependence. }
\label{tab:2}
\end{table}

\section{Axion-photon coupling from the axial Ward identity}
\label{app:derivation_ward_identity}

Here we derive formula~\eqref{eq:fermion1}, 
which forms the basis of the fermionic method.
We start from the axial Ward identity in QCD with background electromagnetic fields for the flavor $f$ in Euclidean space-time,
\begin{equation}\label{eq:awi}
    \partial_{\mu}\langle \bar\psi_f(x)\gamma_5\gamma_\mu\psi_f(x)\rangle^G = 2m_f\langle\bar\psi_f(x)\gamma_5\psi_f(x)\rangle^G + 2\qtop(x) + 2N_c(q_f/e)^2\,\qtop^{\rm EM}(x),
\end{equation}
involving the axial current operator and the pseudoscalar bilinear operator densities. In Eq.~\eqref{eq:awi}, $m_f$ is the mass and $q_f$ the electric charge of the quark and $\langle\cdot\rangle^G$ denotes the expectation value represented by the Grassmannian path integral over quark fields, with the gluonic path integral not yet carried out. In contrast, $\langle\cdot\rangle$ indicates the expectation value over both quark and gluon fields.
Moreover, in Eq.~\eqref{eq:awi}, $N_c=3$ is the number of colors, $\qtop$ is the QCD topological charge density -- appearing as the integrand of Eq.~\eqref{eq:Qtopdef} -- and $\qtop^{\rm EM}$ its electromagnetic counterpart,
\begin{equation}
\label{eq:qtopqtopEMdef}
\qtop(x) = 
\frac{g^2}{64\pi^2}\,\epsilon_{\mu\nu\rho\sigma}G^a_{\mu\nu}(x)G^a_{\rho\sigma}(x)\,,\qquad
    \qtop^{\rm EM} = \frac{e^2}{32\pi^2}\epsilon_{\mu\nu\rho\sigma}F_{\mu\nu}F_{\rho\sigma} = \frac{e^2}{4\pi^2} \,i\bm E\bm B\,.
\end{equation}
$\qtopEM$ is
defined in terms of the electromagnetic field strength tensor $F$, or, equivalently in terms of the electric and magnetic fields. Notice that due to the use of Euclidean space-time, $F$ contains the imaginary electric field components.

Considering a finite system with periodic boundary conditions, such as the one realized by our lattice computations, 
the integral of Eq.~\eqref{eq:awi} over space-time yields,
\begin{equation}
 0 = m_f P_f(\bm E\bm B) + \Qtop + \frac{3 q_f^2}{4\pi^2} \,i\bm E \bm B\cdot V_4\,,
 \label{eq:awi1}
\end{equation}
where $P_f(\bm E\bm B)=\langle \bar\psi_f\gamma_5\psi_f\rangle^G$ is the Grassmannian expectation value of the pseudoscalar operator integrated over space-time, which we already encountered in Eq.~\eqref{eq:trace}.
Since the electromagnetic fields are homogeneous, in the last term the integration merely resulted in the appearance of the space-time volume $V_4=V/T$. In order to make use of Eq.~\eqref{eq:awi1}, we need to carry out the gluonic expectation value as well. For an ensemble generated at nonzero electromagnetic fields, this gives
\begin{equation}\label{eq:qtop_ferm}
     m_f\expv{P_f(\bm E\bm B)}_{\bm E\bm B} = - \expv{\Qtop}_{\bm E\bm B} - \frac{3q_f^2}{4\pi^2}\,i\bm E\bm B\cdot V_4\,.
\end{equation}
Notice that the electromagnetic fields appear in the left hand side in two ways: via the explicit dependence of the operator as well as the implicit dependence through the distribution of gluon fields.
In turn, for an ensemble generated at $\bm E\bm B=0$, the expectation value of the QCD topological charge vanishes, therefore taking the gluonic expectation value of Eq.~\eqref{eq:awi1} in this case gives
\begin{equation}\label{eq:qtop_em_ferm}
     m_f\expv{P_f(\bm E\bm B)}_{0} = -  \frac{3q_f^2}{4\pi^2}\,i\bm E\bm B\cdot V_4\,.
\end{equation}
The last equation provides an estimate for the product $\bm E \bm B$ that fermions on the lattice effectively feel and can therefore be used to partially cancel lattice artifacts.

Combining expressions~\eqref{eq:qtop_ferm} and \eqref{eq:qtop_em_ferm} into the quantity required for the definition~\eqref{eq:apc_def} of the axion-photon coupling, we obtain
\begin{equation}
    \frac{1}{V_4}\frac{\expv{\Qtop}_{\bm E\bm B}}{\bm E \bm B}
    = \frac{3 q_f^2}{4\pi^2}\left[\frac{\expv{P_f(\bm E\bm B)}_{\bm E\bm B}}{\expv{P_f(\bm E\bm B)}_{0}}-1\right].
\end{equation}
leading to Eq.~\eqref{eq:fermion1} of the main text.
Considering axial rotations involving more quark flavors (with nonzero overlap with the singlet transformation) results in linear combinations of Eq.~\eqref{eq:awi} for different $f$. Taking the sum over $N_f$ flavors results in the generalized formula,
\begin{equation}
\frac{1}{V_4}\frac{\expv{\Qtop}_{\bm E\bm B}}{\bm E \bm B}
    %\frac{\gayy f_A}{e^2} + ...
    = \frac{1}{N_f}\sum_{f}\frac{3q_f^2}{4\pi^2}\left[\frac{\sum_{g}m_{g}\expv{P_g(\bm E\bm B)}_{\bm E\bm B}}{\sum_{h}m_{h}\expv{P_h(\bm E\bm B)}_{0}}-1\right].
\end{equation}
For the average of the (mass-degenerate) up and down quarks, this leads to
\begin{equation}
      \gayy=\lim_{\bm E, \bm B\to0} \frac{e^2}{f_A}R^{ud}, \qquad
      R^{ud}=
      %\lim_{\bm E, \bm B\to0} \frac{e^2}{f_A}
      \frac{5}{24\pi^2}\left[\frac{\expv{P_u(\bm E\bm B)}_{\bm E\bm B}+ \expv{P_d(\bm E\bm B)}_{\bm E \bm B}}{\expv{P_u(\bm E\bm B)}_{0}+ \expv{P_d(\bm E\bm B)}_{0}}-1\right]\,,
      \label{eq:fermion2}
\end{equation}

Linear combinations involving both the light and the strange quarks are also possible, but we have observed that those are contaminated by larger lattice artifacts, thus we opted to consider Eqs.~\eqref{eq:fermion1} and~\eqref{eq:fermion2} only.

\section{Computation of the continuum limit of the axion-photon coupling\label{sec:error}}

Here we explain the details of our error analysis, which consists of propagating statistical errors originating from the finite sampling of the lattice configuration ensembles, as well as estimating the systematic errors resulting from the extrapolations of taking $\bm E, \bm B\to0$ and $a\to0$.

Our central observable, which we measured on each ensemble with given $a$, $\bm{E}$ and $\bm{B}$ is the ratio of the topological charge and the CP-odd combination $\bm{EB}$. The ratio is CP-even and Lorentz-invariant, and can therefore be described by the bivariate Taylor expansion,
\begin{equation}
    \label{eq:QEB}
    \frac{1}{V_4}\frac{\expv{\Qtop}_{\bm{E}\bm{B},a}}{\bm{EB}} = 
    \sum_{i,j=0}^{2i+j\le N}
    h_{ij}(a) (\bm{EB})^{2i} (\bm{E}^2+\bm{B}^2)^j\,,
\end{equation}
where $N$ is a non-negative integer so that $2N$ is the highest power of electromagnetic fields appearing in the fit.
The lowest-order coefficient $h_{00}$ is extracted through a series of independent fits of this form at each lattice spacing. This is followed by the continuum extrapolation, which involves -- owing to the scaling properties of our lattice action -- polynomial fits in $a^2$,
\begin{equation}
    \label{eq:gadep}
    h_{00}(a)=\sum_{k=0}^{K} g_k\, a^{2k}\,.
\end{equation}
giving the final value for the axion-photon coupling $\gayy=g_0/f_A$.

To estimate the systematic error of the fits~\eqref{eq:QEB}, i.e.\ of $h_{00}(a)$ at each lattice spacing, we consider
models with different values of $N$,
as well as 
filtered data sets, constructed by excluding all possible combinations of data points and repeating the fits. The maximum value of $N$ for each of these filtered data sets is defined by requiring the number of degrees of freedom of the corresponding fit to be at least equal to one.\footnote{For some of our ensembles, we found the fit direction $\bm E^2+\bm B^2$ to be poorly constrained. To avoid outlier fits, we placed cuts on the magnitude of the leading coefficient $h_{01}(a)$.}
The statistical errors for each of the fits are estimated using the jackknife procedure. 
To combine all of the fits, we associate Gaussian distributions to each fit variant, with means and variances obtained from the jackknife analysis 
and create a single probability distribution by adding them up with a modified version of the Akaike Information Criterion (AIC) as weights, derived in Ref.~\cite{BMW:2014pzb}.
In particular, we use
\begin{equation}
    \text{AIC} \propto \exp{\left[-\frac{1}{2}(\chi^2+2n_{\text{par}}-n_{\text{data}})\right]},
    \label{eq:AIC}
\end{equation}
where $n_{\text{par}}$ is the number of parameters of the fit and $n_{\text{data}}$ the number of data points in the filtered set.
We took the median of the so obtained distribution function as the final result and estimated the combined statistical and systematic error as the symmetric range around the median that contains 68\% of the distribution. For the $\beta=3.95$ ensemble, we have measurements at one value of the electromagnetic fields $\bm E^*$, $\bm B^*$, see Tab.~\ref{tab:1}, giving only one estimate of $h_{00}(a=0.08\textmd{ fm})$. Nevertheless, we associate a systematic error to this result based on the $\beta=3.85$ ensemble. In particular, we take the estimate of $h_{00}(a=0.1\textmd{ fm})$ using only the measurements at $\bm E^*$, $\bm B^*$ and consider its deviation to the final result on that ensemble.

\begin{figure}[htb]
  \centering
  \begin{minipage}[ht]{0.51\textwidth}
    \includegraphics[width=\textwidth]{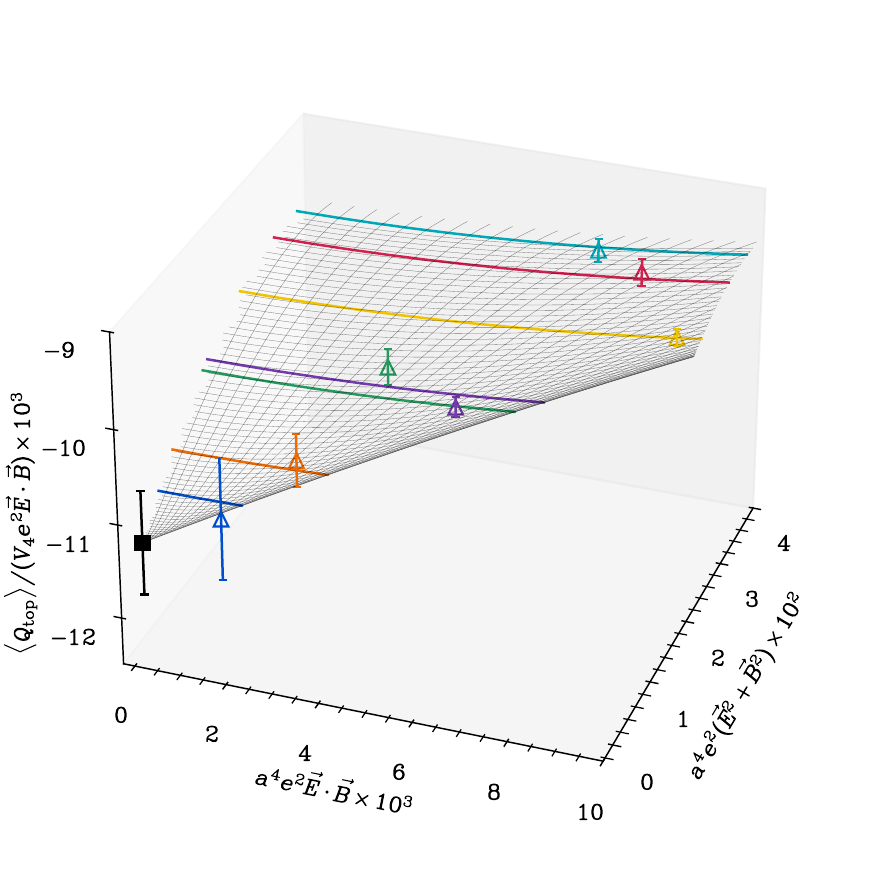}
  \end{minipage}
  \begin{minipage}[ht]{0.48\textwidth}
    \vspace*{.8cm}
    \includegraphics[width=\textwidth]{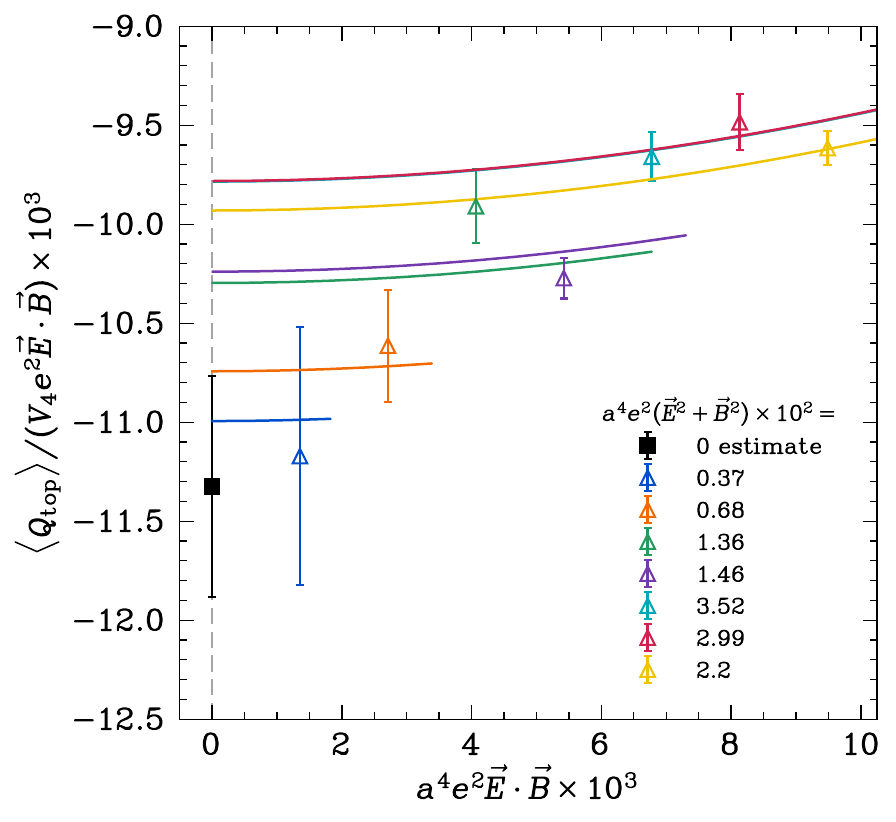}
  \end{minipage}
  \caption{Example fit used for the $\bm{E},\bm{B}\to0$ extrapolation. Different colors indicate different values of $\bm{E}^2+\bm{B}^2$, while the vanishing field estimate is marked by the black filled point. The left panel visualizes the full two-parameter surface, while the right panel shows its constant $\bm{E}^2+\bm{B}^2$ slices corresponding to each data point. For this particular fit, $\chi^2/$d.o.f. $\approx5.37/3$.}
  \label{fig:fits}
\end{figure}

In Fig.~\ref{fig:fits}, we show a representative fit of the type~\eqref{eq:QEB}. In this example, we consider a polynomial with $N=4$ and fit the data obtained in the $a=0.15$~fm  ensemble. The data points are marked by different colors based on their $\bm{E}^2+\bm{B}^2$ value, while the extrapolated result is shown in black. In the left panel, we show the two-parameter surface, with specific constant $\bm{E}^2+\bm{B}^2$ slices corresponding to each data point highlighted. In turn, the right panel shows the highlighted curves and the data points all projected to the same plane.\footnote{Notice that polynomials in the single Lorentz invariant $\bm{EB}$, like the ones used in Ref.~\cite{DElia:2012ifm} for calculating the susceptibility of the QCD vacuum, do not suffice to capture the complete dependence of $\expv{\Qtop}_{\bm{EB}}$ on the electromagnetic fields, even in the weak-field region, as clearly visible in the right panel of Fig.~\ref{fig:fits}.} The surface and the slices are both cut such that $\bm{EB}\leq(\bm{E}^2+\bm{B}^2)/2$, since this holds for any electromagnetic field setup. Error bars of the fitted function are omitted for visibility, but in this example they are the only source for the error on the extrapolated value.

The so obtained distributions are used to draw bootstrap samples in order to propagate errors to the continuum extrapolation~\eqref{eq:gadep}. The latter is carried out in a very similar fashion as the $\bm E, \bm B\to0$ limit,\footnote{We note that we also considered simultaneous trivariate polynomial fits for all of our data points. However, we found that in this setup the final result was highly sensitive to changes in the included data points or in the weighting of the individual fit models. In contrast, the two-step analysis described above provided a robust result.}
using models with $0\le K\leq 3$. In this case, the filtered data sets always included the finest lattice. Moreover, linear fits are excluded when lattice spacings $a>0.2\textmd{ fm}$ are considered.
The final distributions from the different fits are added up using the AIC defined in Eq.~\eqref{eq:AIC}. The final value and the total error is obtained in the same way as in the case of the electromagnetic fits. 
We followed the same strategy for the analysis of the data obtained using the fermionic method as well. 
Example fits for both the gluonic and the fermionic methods are  shown in Fig.~\ref{fig:clim} of the main text.

To further assess the systematics, we considered six different versions of the gluonic method, involving different levels of operator improvement and gradient flow depth as well as a possible rounding of $\Qtop$ to the nearest integer, while for the fermionic method, the four different flavor combinations, as mentioned below Eq.~\eqref{eq:fermion2} already. In each of these cases we obtained compatible results. Since the gluonic definitions are found to be the most accurate, the final continuum extrapolated value and its associated error are defined from the gluonic results. 

The final error of our calculation can be split into a statistical and five different systematical contributions. 
The error associated to the statistical uncertainty was computed by rescaling the variance of the jackknife estimators for each of the final fits, and subtracting the so obtained variance from the total error shown in Fig.~\ref{fig:clim}, following~\cite{Borsanyi:2020mff}. Among the systematical uncertainties, the first one is the error associated to the definition of $\Qtop$. We defined it as the biggest difference of each of the definitions to the central value shown in Fig.~\ref{fig:clim}. The second is the systematic error associated to the continuum limit extrapolations. For this, we considered the difference between the total variance of Fig.~\ref{fig:clim} and the typical variance associated to individual continuum limit fits for a given gluonic definition. 
The remaining part of the total error is the one associated to the $\bm E, \bm B\to0$ extrapolations. To the total variance shown in Fig.~\ref{fig:clim} we added in quadrature the one associated to the volume effects, which we described in App.~\ref{sec:latsetup}. Then, we rescaled our result at the mass-degenerate point with the factor of Eq.~\eqref{eq:factor} and defined its impact on the uncertainty through error propagation.

\begin{figure}[ht]
  \centering
  \includegraphics[width=0.5\textwidth]{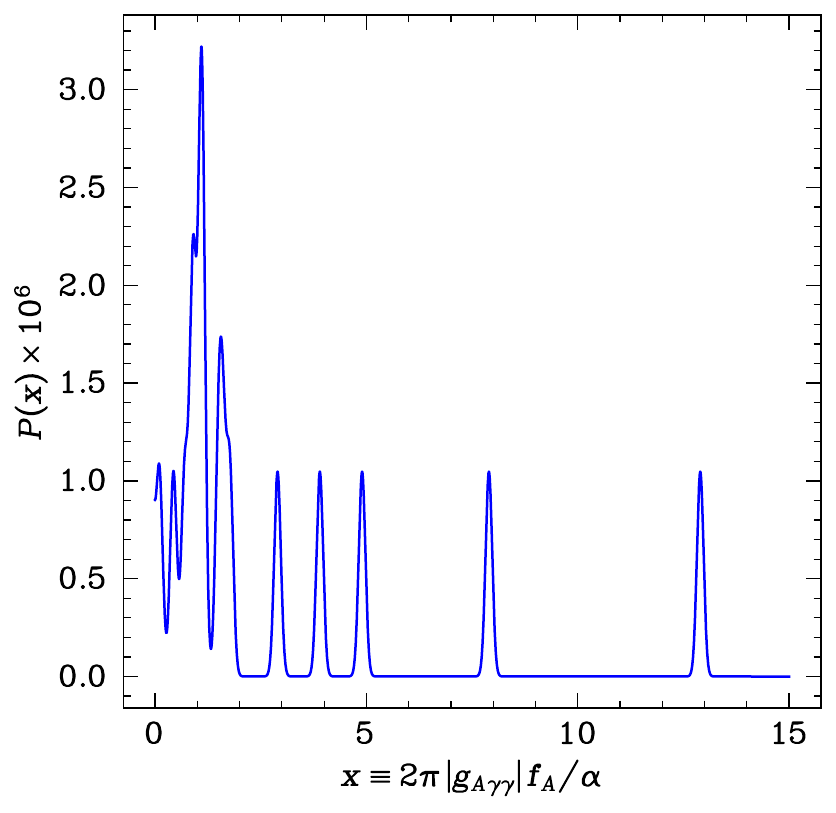}
  \caption{Probability density $P(x)$ of $x\equiv 2\pi|g_{A\gamma\gamma}|f_A/\alpha$ including all possible axion models from Ref.~\cite{DiLuzio:2016sbl}.}
  \label{fig:distro}
\end{figure}

Finally, in Fig.~\ref{fig:distro} we present the distribution of the total coupling, incorporating the probability density of $\gayy$ from our determination as well as the potential axion models from~\cite{DiLuzio:2016sbl}. This distribution is visualized by the color map corresponding to the slopes in Fig.~\ref{axion-photon-full} in the main text.
Notice that to faithfully represent the probability density on the logarithmic scales, the color intensity in  Fig.~\ref{axion-photon-full} indicates $xP(x)$ in order to comply with $P(x)\,\dd x = xP(x)\,\dd (\log x)$.

\bibliography{biblio.bib}

@article{BMW:2014pzb,
    author = "Borsanyi, Sz. and others",
    collaboration = "BMW",
    title = "{Ab initio calculation of the neutron-proton mass difference}",
    eprint = "1406.4088",
    archivePrefix = "arXiv",
    primaryClass = "hep-lat",
    doi = "10.1126/science.1257050",
    journal = "Science",
    volume = "347",
    pages = "1452--1455",
    year = "2015"
}

@article{Gao:2022xqz,
    author = "Gao, Rui and Guo, Zhi-Hui and Oller, J. A. and Zhou, Hai-Qing",
    title = "{Axion-meson mixing in light of recent lattice {\ensuremath{\eta}}{\textendash}{\ensuremath{\eta}}' simulations and their two-photon couplings within U(3) chiral theory}",
    eprint = "2211.02867",
    archivePrefix = "arXiv",
    primaryClass = "hep-ph",
    doi = "10.1007/JHEP04(2023)022",
    journal = "JHEP",
    volume = "04",
    pages = "022",
    year = "2023"
}

@article{Lu:2020rhp,
    author = "Lu, Zhen-Yan and Du, Meng-Lin and Guo, Feng-Kun and Mei{\ss}ner, Ulf-G. and Vonk, Thomas",
    title = "{QCD $\theta$-vacuum energy and axion properties}",
    eprint = "2003.01625",
    archivePrefix = "arXiv",
    primaryClass = "hep-ph",
    doi = "10.1007/JHEP05(2020)001",
    journal = "JHEP",
    volume = "05",
    pages = "001",
    year = "2020"
}

@article{DiLuzio:2016sbl,
    author = "Di Luzio, Luca and Mescia, Federico and Nardi, Enrico",
    title = "{Redefining the Axion Window}",
    eprint = "1610.07593",
    archivePrefix = "arXiv",
    primaryClass = "hep-ph",
    reportNumber = "IPPP-16-99",
    doi = "10.1103/PhysRevLett.118.031801",
    journal = "Phys. Rev. Lett.",
    volume = "118",
    number = "3",
    pages = "031801",
    year = "2017"
}

@article{FlavourLatticeAveragingGroupFLAG:2024oxs,
    author = "Aoki, Y. and others",
    collaboration = "Flavour Lattice Averaging Group (FLAG)",
    title = "{FLAG Review 2024}",
    eprint = "2411.04268",
    archivePrefix = "arXiv",
    primaryClass = "hep-lat",
    reportNumber = "CERN-TH-2024-192, FERMILAB-PUB-24-0785-T",
    month = "11",
    year = "2024"
}

@article{Brandt:2023awt,
    author = "Brandt, Bastian B. and Endr{\H{o}}di, Gergely and Hern{\'a}ndez, Jos{\'e} Javier Hern{\'a}ndez and Mark{\'o}, Gergely and Pannullo, Laurin",
    title = "{Electromagnetic effects on topological observables in QCD}",
    eprint = "2312.14660",
    archivePrefix = "arXiv",
    primaryClass = "hep-lat",
    doi = "10.22323/1.453.0188",
    journal = "PoS",
    volume = "LATTICE2023",
    pages = "188",
    year = "2024"
}

@article{Brandt:2022jfk,
    author = "Brandt, Bastian and Cuteri, Francesca and Endr{\H{o}}di, Gergely and Hern{\'a}ndez, Jos{\'e} Javier Hern{\'a}ndez and Mark{\'o}, Gergely",
    title = "{QCD topology with electromagnetic fields and the axion-photon coupling}",
    eprint = "2212.03385",
    archivePrefix = "arXiv",
    primaryClass = "hep-lat",
    doi = "10.22323/1.430.0174",
    journal = "PoS",
    volume = "LATTICE2022",
    pages = "174",
    year = "2023"
}

@article{Dine:1981rt,
    author = "Dine, Michael and Fischler, Willy and Srednicki, Mark",
    title = "{A Simple Solution to the Strong CP Problem with a Harmless Axion}",
    reportNumber = "Print-81-0320 (IAS,PRINCETON)",
    doi = "10.1016/0370-2693(81)90590-6",
    journal = "Phys. Lett. B",
    volume = "104",
    pages = "199--202",
    year = "1981"
}

@article{Zhitnitsky:1980tq,
    author = "Zhitnitsky, A. R.",
    title = "{On Possible Suppression of the Axion Hadron Interactions. (In Russian)}",
    journal = "Sov. J. Nucl. Phys.",
    volume = "31",
    pages = "260",
    year = "1980"
}

@article{Shifman:1979if,
    author = "Shifman, Mikhail A. and Vainshtein, A. I. and Zakharov, Valentin I.",
    title = "{Can Confinement Ensure Natural CP Invariance of Strong Interactions?}",
    reportNumber = "ITEP-64-1979",
    doi = "10.1016/0550-3213(80)90209-6",
    journal = "Nucl. Phys. B",
    volume = "166",
    pages = "493--506",
    year = "1980"
}

@article{Kim:1979if,
    author = "Kim, Jihn E.",
    title = "{Weak Interaction Singlet and Strong CP Invariance}",
    reportNumber = "UPR-0120T",
    doi = "10.1103/PhysRevLett.43.103",
    journal = "Phys. Rev. Lett.",
    volume = "43",
    pages = "103",
    year = "1979"
}

@article{Borsanyi:2010cj,
    author = "Bors\'anyi, Szabolcs and Endr\H{o}di, Gergely and Fodor, Zoltan and Jakov\'ac, Antal and Katz, Sandor D. and Krieg, Stefan and Ratti, Claudia and Szab\'o, Kalman K.",
    title = "{The QCD equation of state with dynamical quarks}",
    eprint = "1007.2580",
    archivePrefix = "arXiv",
    primaryClass = "hep-lat",
    reportNumber = "WUB-10-12",
    doi = "10.1007/JHEP11(2010)077",
    journal = "JHEP",
    volume = "11",
    pages = "077",
    year = "2010"
}

@article{Bali:2011qj,
    author = "Bali, G. S. and Bruckmann, F. and Endr\H{o}di, G. and Fodor, Z. and Katz, S. D. and Krieg, S. and Sch{\"a}fer, A. and Szab\'o, K. K.",
    title = "{The QCD phase diagram for external magnetic fields}",
    eprint = "1111.4956",
    archivePrefix = "arXiv",
    primaryClass = "hep-lat",
    doi = "10.1007/JHEP02(2012)044",
    journal = "JHEP",
    volume = "02",
    pages = "044",
    year = "2012"
}

@phdthesis{Mazur:2021zgi,
    author = "Mazur, Lukas",
    title = "{Topological Aspects in Lattice QCD}",
    doi = "10.4119/unibi/2956493",
    school = "Bielefeld U.",
    year = "2021"
}

@article{HotQCD:2023ghu,
    author = "Mazur, Lukas and others",
    collaboration = "HotQCD",
    title = "{SIMULATeQCD: A simple multi-GPU lattice code for QCD calculations}",
    eprint = "2306.01098",
    archivePrefix = "arXiv",
    primaryClass = "hep-lat",
    doi = "10.1016/j.cpc.2024.109164",
    journal = "Comput. Phys. Commun.",
    volume = "300",
    pages = "109164",
    year = "2024"
}

@article{Bilson-Thompson:2002xlt,
    author = "Bilson-Thompson, Sundance O. and Leinweber, Derek B. and Williams, Anthony G.",
    title = "{Highly improved lattice field strength tensor}",
    eprint = "hep-lat/0203008",
    archivePrefix = "arXiv",
    reportNumber = "ADP-01-50-T482",
    doi = "10.1016/S0003-4916(03)00009-5",
    journal = "Annals Phys.",
    volume = "304",
    pages = "1--21",
    year = "2003"
}

@article{Luscher:2010iy,
    author = {L\"uscher, Martin},
    title = "{Properties and uses of the Wilson flow in lattice QCD}",
    eprint = "1006.4518",
    archivePrefix = "arXiv",
    primaryClass = "hep-lat",
    reportNumber = "CERN-PH-TH-2010-143",
    doi = "10.1007/JHEP08(2010)071",
    journal = "JHEP",
    volume = "08",
    pages = "071",
    year = "2010",
    note = "[Erratum: JHEP 03, 092 (2014)]"
}

@article{Luscher:2013cpa,
    author = "L{\"u}scher, Martin",
    title = "{Chiral symmetry and the Yang--Mills gradient flow}",
    eprint = "1302.5246",
    archivePrefix = "arXiv",
    primaryClass = "hep-lat",
    reportNumber = "CERN-PH-TH-2013-023",
    doi = "10.1007/JHEP04(2013)123",
    journal = "JHEP",
    volume = "04",
    pages = "123",
    year = "2013"
}

@article{Petreczky:2016vrs,
    author = "Petreczky, Peter and Schadler, Hans-Peter and Sharma, Sayantan",
    title = "{The topological susceptibility in finite temperature QCD and axion cosmology}",
    eprint = "1606.03145",
    archivePrefix = "arXiv",
    primaryClass = "hep-lat",
    doi = "10.1016/j.physletb.2016.09.063",
    journal = "Phys. Lett. B",
    volume = "762",
    pages = "498--505",
    year = "2016"
}

@article{Durr:2013gp,
    author = "D{\"u}rr, Stephan",
    title = "{Taste-split staggered actions: eigenvalues, chiralities and Symanzik improvement}",
    eprint = "1302.0773",
    archivePrefix = "arXiv",
    primaryClass = "hep-lat",
    doi = "10.1103/PhysRevD.87.114501",
    journal = "Phys. Rev. D",
    volume = "87",
    number = "11",
    pages = "114501",
    year = "2013"
}

@article{Borsanyi:2016ksw,
    author = "Bors\'anyi, Sz. and others",
    title = "{Calculation of the axion mass based on high-temperature lattice quantum chromodynamics}",
    eprint = "1606.07494",
    archivePrefix = "arXiv",
    primaryClass = "hep-lat",
    reportNumber = "DESY-16-105",
    doi = "10.1038/nature20115",
    journal = "Nature",
    volume = "539",
    number = "7627",
    pages = "69--71",
    year = "2016"
}

@article{GrillidiCortona:2015jxo,
    author = "Grilli di Cortona, Giovanni and Hardy, Edward and Pardo Vega, Javier and Villadoro, Giovanni",
    title = "{The QCD axion, precisely}",
    eprint = "1511.02867",
    archivePrefix = "arXiv",
    primaryClass = "hep-ph",
    doi = "10.1007/JHEP01(2016)034",
    journal = "JHEP",
    volume = "01",
    pages = "034",
    year = "2016"
}

@misc{AxionLimits,
  author       = {Ciaran O'Hare},
  title        = {cajohare/AxionLimits: AxionLimits},
  month        = jul,
  year         = 2020,
  publisher    = {Zenodo},
  version      = {v1.0},
  doi          = {10.5281/zenodo.3932430},
  howpublished = {\url{https://cajohare.github.io/AxionLimits/}}
}

@article{ParticleDataGroup:2024cfk,
    author = "Navas, S. and others",
    collaboration = "Particle Data Group",
    title = "{Review of particle physics}",
    doi = "10.1103/PhysRevD.110.030001",
    journal = "Phys. Rev. D",
    volume = "110",
    number = "3",
    pages = "030001",
    year = "2024"
}

@article{Endrodi:2024cqn,
    author = "Endr\H{o}di, Gergely",
    title = "{QCD with background electromagnetic fields on the lattice: A review}",
    eprint = "2406.19780",
    archivePrefix = "arXiv",
    primaryClass = "hep-lat",
    doi = "10.1016/j.ppnp.2024.104153",
    journal = "Prog. Part. Nucl. Phys.",
    volume = "141",
    pages = "104153",
    year = "2025"
}

@article{Borsanyi:2020mff,
    author = "Bors\'anyi, Sz. and others",
    title = "{Leading hadronic contribution to the muon magnetic moment from lattice QCD}",
    eprint = "2002.12347",
    archivePrefix = "arXiv",
    primaryClass = "hep-lat",
    doi = "10.1038/s41586-021-03418-1",
    journal = "Nature",
    volume = "593",
    number = "7857",
    pages = "51--55",
    year = "2021"
}

@article{Durr:2005ax,
    author = "D{\"u}rr, Stephan",
    editor = "Michael, Christopher",
    title = "{Theoretical issues with staggered fermion simulations}",
    eprint = "hep-lat/0509026",
    archivePrefix = "arXiv",
    doi = "10.22323/1.020.0021",
    journal = "PoS",
    volume = "LAT2005",
    pages = "021",
    year = "2006"
}

@article{Bali:2012zg,
    author = "Bali, G. S. and Bruckmann, F. and Endr\H{o}di, G. and Fodor, Z. and Katz, S. D. and Sch{\"a}fer, A.",
    title = "{QCD quark condensate in external magnetic fields}",
    eprint = "1206.4205",
    archivePrefix = "arXiv",
    primaryClass = "hep-lat",
    doi = "10.1103/PhysRevD.86.071502",
    journal = "Phys. Rev. D",
    volume = "86",
    pages = "071502",
    year = "2012"
}

@article{Brandt:2024wlw,
    author = "Brandt, Bastian B. and Endr{\H{o}}di, Gergely and Garnacho-Velasco, Eduardo and Mark{\'o}, Gergely",
    title = "{On the absence of the chiral magnetic effect in equilibrium QCD}",
    eprint = "2405.09484",
    archivePrefix = "arXiv",
    primaryClass = "hep-lat",
    doi = "10.1007/JHEP09(2024)092",
    journal = "JHEP",
    volume = "09",
    pages = "092",
    year = "2024"
}

@article{Christenson:1964fg,
    author = "Christenson, J. H. and Cronin, J. W. and Fitch, V. L. and Turlay, R.",
    title = "{Evidence for the $2\pi$ Decay of the $K_2^0$ Meson}",
    doi = "10.1103/PhysRevLett.13.138",
    journal = "Phys. Rev. Lett.",
    volume = "13",
    pages = "138--140",
    year = "1964"
}

@article{Abel:2020pzs,
    author = "Abel, C. and others",
    title = "{Measurement of the Permanent Electric Dipole Moment of the Neutron}",
    eprint = "2001.11966",
    archivePrefix = "arXiv",
    primaryClass = "hep-ex",
    doi = "10.1103/PhysRevLett.124.081803",
    journal = "Phys. Rev. Lett.",
    volume = "124",
    number = "8",
    pages = "081803",
    year = "2020"
}

@article{Peccei:1977hh,
    author = "Peccei, R. D. and Quinn, Helen R.",
    title = "{CP Conservation in the Presence of Instantons}",
    reportNumber = "ITP-568-STANFORD",
    doi = "10.1103/PhysRevLett.38.1440",
    journal = "Phys. Rev. Lett.",
    volume = "38",
    pages = "1440--1443",
    year = "1977"
}

@article{Peccei:1977ur,
    author = "Peccei, R. D. and Quinn, Helen R.",
    title = "{Constraints Imposed by CP Conservation in the Presence of Instantons}",
    reportNumber = "ITP-572-STANFORD",
    doi = "10.1103/PhysRevD.16.1791",
    journal = "Phys. Rev. D",
    volume = "16",
    pages = "1791--1797",
    year = "1977"
}

@article{DiLuzio:2020wdo,
    author = "Di Luzio, Luca and Giannotti, Maurizio and Nardi, Enrico and Visinelli, Luca",
    title = "{The landscape of QCD axion models}",
    eprint = "2003.01100",
    archivePrefix = "arXiv",
    primaryClass = "hep-ph",
    reportNumber = "DESY 20-036, DESY-20-036",
    doi = "10.1016/j.physrep.2020.06.002",
    journal = "Phys. Rept.",
    volume = "870",
    pages = "1--117",
    year = "2020"
}

@article{Brandt:2024gso,
    author = "Brandt, B. B. and Endr{\H{o}}di, G. and Hern{\'a}ndez, J. J. Hern{\'a}ndez and Mark{\'o}, G.",
    title = "{Impact of extreme magnetic fields on the QCD topological susceptibility in the vicinity of the crossover region}",
    eprint = "2409.00796",
    archivePrefix = "arXiv",
    primaryClass = "hep-lat",
    doi = "10.1007/JHEP12(2024)228",
    journal = "JHEP",
    volume = "12",
    pages = "228",
    year = "2025",
    note = "[Erratum: JHEP 03, 034 (2025)]"
}

@article{Bonanno:2019xhg,
    author = "Bonanno, Claudio and Clemente, Giuseppe and D'Elia, Massimo and Sanfilippo, Francesco",
    title = "{Topology via spectral projectors with staggered fermions}",
    eprint = "1908.11832",
    archivePrefix = "arXiv",
    primaryClass = "hep-lat",
    doi = "10.1007/JHEP10(2019)187",
    journal = "JHEP",
    volume = "10",
    pages = "187",
    year = "2019"
}

@article{DElia:2012ifm,
    author = "D'Elia, Massimo and Mariti, Marco and Negro, Francesco",
    title = "{Susceptibility of the QCD vacuum to CP-odd electromagnetic background fields}",
    eprint = "1209.0722",
    archivePrefix = "arXiv",
    primaryClass = "hep-lat",
    reportNumber = "IFUP-TH-2012-16",
    doi = "10.1103/PhysRevLett.110.082002",
    journal = "Phys. Rev. Lett.",
    volume = "110",
    number = "8",
    pages = "082002",
    year = "2013"
}

@article{Meggiolaro:2025yiu,
    author = "Meggiolaro, Enrico and Tamburini, Mirko",
    title = "{New study of the interactions of the axion with mesons and photons using a chiral effective Lagrangian model}",
    eprint = "2502.13615",
    archivePrefix = "arXiv",
    primaryClass = "hep-ph",
    reportNumber = "IFUP-TH/2025",
    doi = "10.1103/PhysRevD.111.095024",
    journal = "Phys. Rev. D",
    volume = "111",
    number = "9",
    pages = "095024",
    year = "2025"
}

@article{Atiyah:1968mp,
    author = "Atiyah, M. F. and Singer, I. M.",
    title = "{The Index of elliptic operators. 1}",
    doi = "10.2307/1970715",
    journal = "Annals Math.",
    volume = "87",
    pages = "484--530",
    year = "1968"
}

@article{Athenodorou:2022aay,
    author = "Athenodorou, Andreas and Bonanno, Claudio and Bonati, Claudio and Clemente, Giuseppe and D'Angelo, Francesco and D'Elia, Massimo and Maio, Lorenzo and Martinelli, Guido and Sanfilippo, Francesco and Todaro, Antonino",
    title = "{Topological susceptibility of N$_{f}$ = 2 + 1 QCD from staggered fermions spectral projectors at high temperatures}",
    eprint = "2208.08921",
    archivePrefix = "arXiv",
    primaryClass = "hep-lat",
    doi = "10.1007/JHEP10(2022)197",
    journal = "JHEP",
    volume = "10",
    pages = "197",
    year = "2022"
}

@article{Bonati:2015vqz,
    author = "Bonati, Claudio and D'Elia, Massimo and Mariti, Marco and Martinelli, Guido and Mesiti, Michele and Negro, Francesco and Sanfilippo, Francesco and Villadoro, Giovanni",
    title = "{Axion phenomenology and $\theta$-dependence from $N_f = 2+1$ lattice QCD}",
    eprint = "1512.06746",
    archivePrefix = "arXiv",
    primaryClass = "hep-lat",
    reportNumber = "IFUP-TH-2015-15",
    doi = "10.1007/JHEP03(2016)155",
    journal = "JHEP",
    volume = "03",
    pages = "155",
    year = "2016"
}

@article{Taniguchi:2016tjc,
    author = "Taniguchi, Yusuke and Kanaya, Kazuyuki and Suzuki, Hiroshi and Umeda, Takashi",
    title = "{Topological susceptibility in finite temperature ( 2+1 )-flavor QCD using gradient flow}",
    eprint = "1611.02411",
    archivePrefix = "arXiv",
    primaryClass = "hep-lat",
    reportNumber = "UTHEP-697, UTCCS-P-93, KYUSHU-HET-172",
    doi = "10.1103/PhysRevD.95.054502",
    journal = "Phys. Rev. D",
    volume = "95",
    number = "5",
    pages = "054502",
    year = "2017"
}

@article{Rubin:1970zza,
    author = "Rubin, Vera C. and Ford, Jr., W. Kent",
    title = "{Rotation of the Andromeda Nebula from a Spectroscopic Survey of Emission Regions}",
    doi = "10.1086/150317",
    journal = "Astrophys. J.",
    volume = "159",
    pages = "379--403",
    year = "1970"
}

@article{Zwicky:1933gu,
    author = "Zwicky, F.",
    title = "{Die Rotverschiebung von extragalaktischen Nebeln}",
    doi = "10.1007/s10714-008-0707-4",
    journal = "Helv. Phys. Acta",
    volume = "6",
    pages = "110--127",
    year = "1933"
}

@article{Clowe:2006eq,
    author = "Clowe, Douglas and Bradac, Marusa and Gonzalez, Anthony H. and Markevitch, Maxim and Randall, Scott W. and Jones, Christine and Zaritsky, Dennis",
    title = "{A direct empirical proof of the existence of dark matter}",
    eprint = "astro-ph/0608407",
    archivePrefix = "arXiv",
    reportNumber = "SLAC-PUB-12078",
    doi = "10.1086/508162",
    journal = "Astrophys. J. Lett.",
    volume = "648",
    pages = "L109--L113",
    year = "2006"
}

@article{Cirelli:2024ssz,
    author = "Cirelli, Marco and Strumia, Alessandro and Zupan, Jure",
    title = "{Dark Matter}",
    eprint = "2406.01705",
    archivePrefix = "arXiv",
    primaryClass = "hep-ph",
    month = "6",
    year = "2024"
}

@article{Ringwald:2024uds,
    author = "Ringwald, Andreas",
    title = "{Review on Axions}",
    eprint = "2404.09036",
    archivePrefix = "arXiv",
    primaryClass = "hep-ph",
    reportNumber = "DESY-24-054",
    doi = "10.1088/1742-6596/3162/1/012006",
    journal = "J. Phys. Conf. Ser.",
    volume = "3162",
    number = "1",
    pages = "012006",
    year = "2025"
}

@article{IAXO:2020wwp,
    author = "Abeln, A. and others",
    collaboration = "IAXO",
    title = "{Conceptual design of BabyIAXO, the intermediate stage towards the International Axion Observatory}",
    eprint = "2010.12076",
    archivePrefix = "arXiv",
    primaryClass = "physics.ins-det",
    doi = "10.1007/JHEP05(2021)137",
    journal = "JHEP",
    volume = "05",
    pages = "137",
    year = "2021"
}

@article{Stern:2016bbw,
    author = "Stern, I.",
    title = "{ADMX Status}",
    eprint = "1612.08296",
    archivePrefix = "arXiv",
    primaryClass = "physics.ins-det",
    doi = "10.22323/1.282.0198",
    journal = "PoS",
    volume = "ICHEP2016",
    pages = "198",
    year = "2016"
}

@article{Alesini:2023qed,
    author = "Alesini, David and others",
    title = "{The future search for low-frequency axions and new physics with the FLASH resonant cavity experiment at Frascati National Laboratories}",
    eprint = "2309.00351",
    archivePrefix = "arXiv",
    primaryClass = "physics.ins-det",
    reportNumber = "CA21106; CA21136",
    doi = "10.1016/j.dark.2023.101370",
    journal = "Phys. Dark Univ.",
    volume = "42",
    pages = "101370",
    year = "2023"
}

@article{Bajjali:2023uis,
    author = "Bajjali, Fayez and others",
    title = "{First results from BRASS-p broadband searches for hidden photon dark matter}",
    eprint = "2306.05934",
    archivePrefix = "arXiv",
    primaryClass = "hep-ex",
    doi = "10.1088/1475-7516/2023/08/077",
    journal = "JCAP",
    volume = "08",
    pages = "077",
    year = "2023"
}

@article{BREAD:2021tpx,
    author = "Liu, Jesse and others",
    collaboration = "BREAD",
    title = "{Broadband Solenoidal Haloscope for Terahertz Axion Detection}",
    eprint = "2111.12103",
    archivePrefix = "arXiv",
    primaryClass = "physics.ins-det",
    reportNumber = "FERMILAB-PUB-21-694-AD-PPD-TD",
    doi = "10.1103/PhysRevLett.128.131801",
    journal = "Phys. Rev. Lett.",
    volume = "128",
    number = "13",
    pages = "131801",
    year = "2022"
}

@article{Beurthey:2020yuq,
    author = "Beurthey, S. and others",
    title = "{MADMAX Status Report}",
    eprint = "2003.10894",
    archivePrefix = "arXiv",
    primaryClass = "physics.ins-det",
    month = "3",
    year = "2020"
}

@article{ALPHA:2022rxj,
    author = "Millar, Alexander J. and others",
    collaboration = "ALPHA",
    title = "{Searching for dark matter with plasma haloscopes}",
    eprint = "2210.00017",
    archivePrefix = "arXiv",
    primaryClass = "hep-ph",
    reportNumber = "FERMILAB-PUB-22-739-T",
    doi = "10.1103/PhysRevD.107.055013",
    journal = "Phys. Rev. D",
    volume = "107",
    number = "5",
    pages = "055013",
    year = "2023"
}

@article{Ouellet:2018beu,
    author = "Ouellet, Jonathan L. and others",
    title = "{First Results from ABRACADABRA-10 cm: A Search for Sub-$\mu$eV Axion Dark Matter}",
    eprint = "1810.12257",
    archivePrefix = "arXiv",
    primaryClass = "hep-ex",
    doi = "10.1103/PhysRevLett.122.121802",
    journal = "Phys. Rev. Lett.",
    volume = "122",
    number = "12",
    pages = "121802",
    year = "2019"
}

@article{Salemi:2021gck,
    author = "Salemi, Chiara P. and others",
    title = "{Search for Low-Mass Axion Dark Matter with ABRACADABRA-10~cm}",
    eprint = "2102.06722",
    archivePrefix = "arXiv",
    primaryClass = "hep-ex",
    doi = "10.1103/PhysRevLett.127.081801",
    journal = "Phys. Rev. Lett.",
    volume = "127",
    number = "8",
    pages = "081801",
    year = "2021"
}

@article{Crisosto:2019fcj,
    author = "Crisosto, N. and Sikivie, P. and Sullivan, N. S. and Tanner, D. B. and Yang, J. and Rybka, G.",
    title = "{ADMX SLIC: Results from a Superconducting $LC$ Circuit Investigating Cold Axions}",
    eprint = "1911.05772",
    archivePrefix = "arXiv",
    primaryClass = "astro-ph.CO",
    doi = "10.1103/PhysRevLett.124.241101",
    journal = "Phys. Rev. Lett.",
    volume = "124",
    number = "24",
    pages = "241101",
    year = "2020"
}

@article{Gramolin:2020ict,
    author = "Gramolin, Alexander V. and Aybas, Deniz and Johnson, Dorian and Adam, Janos and Sushkov, Alexander O.",
    title = "{Search for axion-like dark matter with ferromagnets}",
    eprint = "2003.03348",
    archivePrefix = "arXiv",
    primaryClass = "hep-ex",
    doi = "10.1038/s41567-020-1006-6",
    journal = "Nature Phys.",
    volume = "17",
    number = "1",
    pages = "79--84",
    year = "2021"
}

@article{Zhang:2021bpa,
    author = "Zhang, Zhongyue and Horns, Dieter and Ghosh, Oindrila",
    title = "{Search for dark matter with an LC circuit}",
    eprint = "2111.04541",
    archivePrefix = "arXiv",
    primaryClass = "hep-ex",
    doi = "10.1103/PhysRevD.106.023003",
    journal = "Phys. Rev. D",
    volume = "106",
    number = "2",
    pages = "023003",
    year = "2022"
}

@article{DMRadio:2022pkf,
    author = "Brouwer, L. and others",
    collaboration = "DMRadio",
    title = "{Projected sensitivity of DMRadio-m3: A search for the QCD axion below 1{\,}{\,}{\ensuremath{\mu}}eV}",
    eprint = "2204.13781",
    archivePrefix = "arXiv",
    primaryClass = "hep-ex",
    doi = "10.1103/PhysRevD.106.103008",
    journal = "Phys. Rev. D",
    volume = "106",
    number = "10",
    pages = "103008",
    year = "2022"
}

@article{Aybas:2021nvn,
    author = "Aybas, Deniz and others",
    title = "{Search for Axionlike Dark Matter Using Solid-State Nuclear Magnetic Resonance}",
    eprint = "2101.01241",
    archivePrefix = "arXiv",
    primaryClass = "hep-ex",
    doi = "10.1103/PhysRevLett.126.141802",
    journal = "Phys. Rev. Lett.",
    volume = "126",
    number = "14",
    pages = "141802",
    year = "2021"
}

@article{Bauer:2020jbp,
    author = "Bauer, Martin and Neubert, Matthias and Renner, Sophie and Schnubel, Marvin and Thamm, Andrea",
    title = "{The Low-Energy Effective Theory of Axions and ALPs}",
    eprint = "2012.12272",
    archivePrefix = "arXiv",
    primaryClass = "hep-ph",
    reportNumber = "IPPP/20/69, MITP/20-070 SISSA 30/2020/FISI, ZH-TH-47/20",
    doi = "10.1007/JHEP04(2021)063",
    journal = "JHEP",
    volume = "04",
    pages = "063",
    year = "2021"
}

@article{Choi:2021kuy,
    author = "Choi, Kiwoon and Im, Sang Hui and Kim, Hee Jung and Seong, Hyeonseok",
    title = "{Precision axion physics with running axion couplings}",
    eprint = "2106.05816",
    archivePrefix = "arXiv",
    primaryClass = "hep-ph",
    reportNumber = "CTPU-PTC-21-26",
    doi = "10.1007/JHEP08(2021)058",
    journal = "JHEP",
    volume = "08",
    pages = "058",
    year = "2021"
}

@article{DiLuzio:2023tqe,
    author = "Di Luzio, Luca and Giannotti, Maurizio and Mescia, Federico and Nardi, Enrico and Okawa, Shohei and Piazza, Gioacchino",
    title = "{Running effects on QCD axion phenomenology}",
    eprint = "2305.11958",
    archivePrefix = "arXiv",
    primaryClass = "hep-ph",
    doi = "10.1103/PhysRevD.108.115004",
    journal = "Phys. Rev. D",
    volume = "108",
    number = "11",
    pages = "115004",
    year = "2023"
}

@article{Bauer:2017ris,
    author = "Bauer, Martin and Neubert, Matthias and Thamm, Andrea",
    title = "{Collider Probes of Axion-Like Particles}",
    eprint = "1708.00443",
    archivePrefix = "arXiv",
    primaryClass = "hep-ph",
    reportNumber = "MITP-17-047",
    doi = "10.1007/JHEP12(2017)044",
    journal = "JHEP",
    volume = "12",
    pages = "044",
    year = "2017"
}

\end{document}